\documentclass[12pt,preprint]{aastex}
\def\lsim{\raise0.3ex\hbox{$<$}\kern-0.75em{\lower0.65ex\hbox{$\sim$}}}
\def\gsim{\raise0.3ex\hbox{$>$}\kern-0.75em{\lower0.65ex\hbox{$\sim$}}}

\begin{document}
\title{Magnetically Regulated Star Formation in 3D: The Case of 
Taurus Molecular Cloud Complex} 
 
\author{Fumitaka Nakamura\altaffilmark{1} and Zhi-Yun Li\altaffilmark{2}} 
\altaffiltext{1}{Faculty of Education, Niigata University,
8050 Ikarashi-2, Niigata 950-2181, Japan; fnakamur@ed.niigata-u.ac.jp}
\altaffiltext{2}{Department of Astronomy, University of Virginia,
P. O. Box 400325, Charlottesville, VA 22904; zl4h@virginia.edu}

\begin{abstract}
We carry out three-dimensional MHD simulations of star formation
in turbulent, magnetized clouds, including ambipolar diffusion
and feedback from protostellar outflows. The calculations focus
on relatively diffuse clouds threaded by a strong magnetic field 
capable of resisting severe tangling by turbulent motions and 
retarding global gravitational contraction in the cross-field 
direction. They are motivated by observations of the Taurus 
molecular cloud complex (and, to a lesser extent, Pipe Nebula), 
which shows an ordered large-scale magnetic field, as 
well as elongated condensations that are generally perpendicular 
to the large-scale field. We find that stars form in earnest in 
such clouds when enough material has 
settled gravitationally along the field lines that the 
mass-to-flux ratios of the condensations approach the critical 
value. Only a small fraction (of order $1\%$ or less) of the 
nearly magnetically-critical, 
condensed material is turned into 
stars per local free-fall time, however. The slow star formation 
takes place in condensations that are moderately supersonic; 
it is regulated primarily by magnetic fields, rather than turbulence. 
The quiescent condensations are surrounded by diffuse halos 
that are much more turbulent, as observed in the Taurus complex. 
Strong support for magnetic regulation of star formation in 
this complex comes from the extremely slow conversion of the 
already condensed, relatively quiescent C$^{18}$O gas into 
stars, at a rate two orders of magnitude below the maximum, 
free-fall value.
We analyze the properties of dense 
cores, including their mass spectrum, which resembles the 
stellar initial mass function. 
\end{abstract}


\section{Introduction}
\label{intro}

Stars form in molecular clouds that are both turbulent and magnetized. 
The relative importance of the turbulence and magnetic field in 
controlling star formation is a matter of debate (McKee \& Ostriker 
2007). Early quantitative studies have concentrated on the formation 
and evolution of individual (low-mass) cores out of quiescent, 
magnetically supported clouds (Nakano 1984; Shu et al. 1987; 
Mouschovias \& Ciolek 1999). Recent numerical simulations have 
focused more on the role of turbulence in the dynamics of larger 
clouds and core formation in them (as reviewed in, e.g., Mac Low 
\& Klessen 2004). Ultimately, the debate must be settled by 
observations, especially direct measurements of the magnetic 
field strength in the bulk of the molecular gas, which are 
generally difficult to do (e.g., Heiles \& Crutcher 2005). In the 
absence of such measurements, we have to rely on indirect evidence.

The best observed region of active star formation is arguably 
the nearby Taurus molecular cloud complex. Although it is not 
clear whether the complex is magnetically supported or not as 
a whole, the more diffuse regions are probably magnetically 
dominated. The strongest evidence comes from thin strands of 
$^{12}$CO emission that are aligned with the directions of 
the local magnetic field (Heyer et al. 2008), as traced by 
polarization of background 
star light. The magnetic field in the strands is apparently
strong enough to induce a measurable difference between the
turbulent velocities along and perpendicular to the field
direction. Heyer et al. concluded that the field strength 
is probably high enough to render the relatively diffuse 
striated region subcritical, with a magnetic flux-to-mass 
ratio greater than the critical value $2\pi G^{1/2}$ (Nakano 
\& Nakamura 1978). The CO striations are strikingly 
similar to those observed in the nearby Riegel-Crutcher HI 
cloud, mapped recently by McClure-Griffiths et al. (2006), 
using 21cm absorption against the strong continuum emission
towards the galactic center region. Its hair-like strands 
are also along the directions of the local magnetic field, 
again traced by the polarization vectors of the background 
star light. The authors estimated a field strength of $\sim 
30$~$\mu$G inside the strands. Kazes \& Crutcher (1986) 
measured the line-of-sight field strength at a nearby 
location, and found $B_{los}\sim 18~\mu$G, which is 
consistent with the above estimate if a correction of $\sim 
2$ is applied for projection effect. These are clear examples 
of magnetic domination in, respectively, cold HI
% (see also Orion Veil, Brogan et al.??) 
and relatively diffuse molecular clouds. 

It is unclear, however, how representative these two clouds are. 
In particular, the degree of magnetization is uncertain in 
giant molecular clouds (GMCs), where the majority of stars 
form. Elmegreen (2007) argued that the envelopes of GMCs, where 
most of the molecular gas resides, are probably magnetically 
critical or even subcritical, and can be long-lived if not for 
disruption by rapid star formation in GMC cores, which are 
probably supercritical, with internal dynamics dominated by 
turbulence; this supposition needs to be tested by magnetic 
field observations on GMC scales (e.g., Novak, et al. 2007). 
Nevertheless, the 
Taurus cloud complex and the Riegel-Crutcher cloud are probably  
best mapped cloud in CO and HI, respectively. How such relatively 
diffuse, apparently magnetically dominated clouds proceed to 
condense and collapse to form stars is the focus of our paper. 

Diffuse clouds are expected to condense more easily along than 
across the field lines as long as they are dominated by ordered 
large-scale magnetic fields. Such an anisotropic condensation 
can increase the density greatly with little corresponding 
increase in the magnetic field strength. This expectation is 
consistent with the available Zeeman measurements of field 
strength, which is more or less constant below a number density 
of order $10^3$~cm$^{-3}$ (Heiles \& Crutcher 2005). Above this 
density, the field strength tends to increase with density, indicating 
that contraction perpendicular to the field lines becomes 
significant. We interpret the break in the observed field 
strength-density relation as marking the point where the 
self-gravity of the cloud becomes dynamically important in the 
cross-field direction. Before this point is reached, one expects 
the self-gravity to first become significant along the field 
lines (where there is no magnetic support), pulling matter into 
condensations, which can be either a sheet, a filament, 
or even a knot, depending on the degree of anisotropy in the 
initial mass distribution. 

In the case of the Taurus molecular cloud complex, most of the 
dense molecular gas traced by $^{13}$CO and especially C$^{18}$O 
are distributed
in structures elongated more or less perpendicular to the 
large-scale magnetic field (e.g., Onishi et 
al. 1996, 2002; Goldsmith et al. 2008). It has long been suspected 
that they have condensed along the field lines (e.g., Heyer et 
al. 1987; Tamura et al. 1987). Cross-field contraction must have 
already taken place in the condensed structures, at least locally, 
since stars have been forming in these structures for at least
a few million years (Kenyon \& Hartmann 1995). Palla \& Stahler
(2002) examined the star formation pattern in the Taurus
clouds in both space and time, concluding that stars began 
to form at a low level at least 10 million years ago, in 
a spatially dispersed fashion. The majority of the stars are
formed, however, in the last 3 million years, near where the 
dense gas is observed today. This pattern of accelerating 
star formation is precisely what is expected of a magnetically
dominated cloud that condenses in two stages: first along the
field line, when little star formation is expected, except 
perhaps in some pockets that have an exceptionally weak local 
magnetic field to begin with or have their magnetic fluxes 
reduced by an exceptionally strong compression-enhanced ambipolar 
diffusion, and then across the field, when more active star
formation can take place throughout the dense sheets and 
filaments that have condensed along the field lines. A strong 
support for this supposition comes from the fact that star 
formation is slow even in dense gas: the rate of star 
formation for the last 3 million years is about ${\dot M}_*
\sim 5\times 10^{-5}~M_\odot$yr$^{-1}$ (Goldsmith et al. 2008), 
two orders of magnitude below the free-fall rate for the dense 
($\sim 10^4$~cm$^{-3}$) C$^{18}$O gas (Onishi et al. 2002; see 
\S~\ref{taurus} for actual numbers); Krumholz \& Tan (2006) found 
that a similar result holds for objects of a wide range of density. 
Although the observed moderately supersonic motions (${\cal M}\sim 
2$) can provide some support for the dense gas, we believe that 
the star formation is slow mainly because the bulk of the dense 
gas remains magnetically supported even after cross-field 
contraction has begun in localized regions in the condensed 
structures. 
This requirement can naturally be satisfied if such structures 
are marginally magnetically critical (Basu \& Ciolek 2000), 
where individual stars can form on a relatively short time 
scale, but the overall rate of star formation remains well 
below the free-fall value, as we have demonstrated using 
simulations in 2D sheet-like geometry (Li \& Nakamura 2004 
and Nakamura \& Li 2005; LN04 and NL05 hereafter; see also 
Kudoh et al. 2007 and Kudoh \& Basu 2008). The Taurus cloud 
complex is discussed further in \S~\ref{taurus}, along 
with the Pipe Nebula (\S~\ref{pipe}), which may represent an 
earlier phase of star formation in a magnetically dominated 
cloud (Lada et al. 2008; F. Alves et al., in preparation).   

The rest of the paper is organized as follows. Section~\ref{setup} 
describes the model formulation, including governing equations, 
initial and boundary conditions, and numerical code. It is followed
by three sections on numerical results. Section~\ref{result} concentrates 
on a standard model with a particular combination of initial turbulence 
and outflow feedback. It illustrates the essential features of the 
magnetically regulated star formation in three dimensions, including 
strong stratifications in density and turbulent speed, broad 
probability distribution functions (PDFs) of volume and column 
densities and, most importantly, low rate of star formation. This 
model is contrasted, in \S~\ref{other}, with others that have 
different levels of initial turbulence and outflow feedback. We 
devote \S~\ref{core} to a detailed analysis of dense cores, including
their shapes, mass spectrum, velocity dispersions, flux-to-mass
ratios, angular momenta, and whether they are bound or not. In 
\S~\ref{discussion}, we outline a general scenario of magnetically
regulated star formation in quiescent condensations of relatively
diffuse, turbulent clouds, and make connection to observations 
of the Taurus molecular cloud complex and Pipe Nebula. The main 
results of the paper are summarized in \S~\ref{conclude}. Readers 
interested only in the general scenario and their connection to 
observations can skip to \S~\ref{discussion}.

\section{Model Formulation}
\label{setup}

\subsection{Governing Equations}

The basic equations that govern the evolution of isothermal, weakly 
ionized, magnetized molecular gas are (see, e.g., Shu 1991)
\begin{equation}
{\partial \rho_n \over \partial t} + \nabla \cdot (\rho_n
 \mbox{\boldmath{$V$}}_n) = 0 \ ,
\end{equation}
\begin{equation}
\rho_n {\partial \mbox{\boldmath{$V$}}_n \over \partial t} +
\rho_n (\mbox{\boldmath{$V$}}_n \cdot \nabla \mbox{\boldmath{$V$}}_n)
= -\rho_n \nabla \Psi - c_s^2 \nabla \rho_n + {1 \over 4 \pi}
(\nabla \times \mbox{\boldmath{$B$}}) \times \mbox{\boldmath{$B$}} \ , 
\end{equation}
\begin{equation}
{\partial \mbox{\boldmath{$B$}} \over \partial t}
=\nabla \times (\mbox{\boldmath{$V$}}_i \times \mbox{\boldmath{$B$}}) \ ,
\end{equation}
\begin{equation}
\nabla ^2 \Psi = 4 \pi G \rho_n  \ , 
\end{equation}
where 
$\rho$, $\mbox{\boldmath{$V$}}$, $\mbox{\boldmath{$B$}}$, $c_s$,
and $\Psi$ are, respectively, 
density, velocity, magnetic field, isothermal sound speed, 
and gravitational potential, with the subscripts $n$ and $i$ 
denoting neutrals and ions. The drift velocity between ion 
and neutral is 
\begin{equation}
\mbox{\boldmath{$V$}}_i - \mbox{\boldmath{$V$}}_n = {t_c \over 4 \pi
 \rho_n} (\nabla \times \mbox{\boldmath{$B$}}) \times 
\mbox{\boldmath{$B$}}  \ , 
\end{equation}
where the coupling time between the magnetic field and neutral matter,
$t_c$, is given by 
\begin{equation}
t_c = {1.4 \over \gamma C \rho_n^{1/2}},
\end{equation}
in the simplest case where the coupling is provided by ions
that are well tied to the field lines and the ion density $\rho_i$
is related to the neutral density by the canonical expression
$\rho_i = C \rho_n^{1/2}$. We adopt an ion-neutral drag coefficient 
$\gamma = 3.5\times 10^{13}$~cm$^3$~g$^{-1}$~s$^{-1}$ and $C = 3 
\times 10^{-16}$~cm$^{-3/2}$~g$^{1/2}$ (Shu 1991). The factor 1.4 
in the above equation comes from the fact that the cross section 
for ion-helium collision is small compared to that of ion-hydrogen 
collision (Mouschovias \& Morton 1991). These governing equations 
are solved numerically subject to a set of initial and boundary 
conditions. 

\subsection{Initial and Boundary Conditions}

Our simulations are carried out in a cubic box (of size $L$), with
standard periodic conditions imposed at the boundaries. The cloud 
is assumed to have a uniform density $\rho_0$ initially. The 
corresponding Jeans length is 
\begin{equation} 
L_J= \left({\pi c_s^2/G \rho_0}\right)^{1/2}, 
\end{equation}
where the isothermal sound speed $c_s=1.88\times 10^4(T/10 
K)^{1/2}$~cm/s, with $T$ being the cloud temperature. The 
initial density $\rho_0 = 4.68\times 10^{-24} n_{H_2,0}$~g~cm$^{-3}$, 
where $n_{H_2,0}$ is the initial number density of molecular 
hydrogen, assuming 1 He for every 10 H atoms. 

We consider relatively diffuse molecular clouds, with a fiducial 
H$_2$ number density of $250$~cm$^{-3}$. It is the value adopted 
by Heyer et al. (2008) for the sub-region in Taurus that shows 
magnetically aligned $^{12}$CO striations. Scaling the density by 
this value, we have 
\begin{equation}
L_J=1.22 \left(\frac{T}{10 K}\right)^{1/2} 
\left(\frac{n_{H_2,0}}{250\ {\rm cm}^{-3}}\right)^{-1/2} {\rm pc}, 
\end{equation}
and a Jeans mass 
\begin{equation} 
M_J= \rho_0 L_J^3 = 31.6 M_\odot \left(\frac{T}{10 K}\right)^{3/2} 
\left(\frac{n_{H_2,0}}{250\ {\rm cm}^{-3}}\right)^{-1/2}.
\end{equation} 
The Jeans length is smaller than the dimensions of molecular cloud 
complexes such as the Taurus clouds, which are typically tens of 
parsecs. Although global simulations on the complex scale are
desirable, they are prohibitively expensive if individual star 
formation events are to be resolved at the same time. After some 
experimentation, we settled on a moderate size for the simulation 
box, with an initial Jeans number $n_J=L/L_J=2$. It represents 
a relatively small piece of a larger complex. Scaling the Jeans 
number by 2, we have a box size 
\begin{equation}
L=2.44 \left(\frac{n_J}{2}\right) \left(\frac{T}{10 K}\right)^{1/2} 
\left(\frac{n_{H_2,0}}{250\ {\rm cm}^{-3}}\right)^{-1/2} {\rm pc}, 
\end{equation}
and a total mass 
\begin{equation} 
M_{tot}= 253\ M_\odot \left(\frac{n_J}{2}\right)^3 \left(\frac{T}{10 
K}\right)^{3/2}\left(\frac{n_{H_2,0}}{250\ {\rm cm}^{-3}}\right)^{-1/2}
\end{equation} 
in the computation domain. Even though the initial Jeans number
adopted is relatively small, there is enough material inside 
the box to form many low-mass stars, especially in condensations 
where the Jeans mass is much reduced. 

We impose a uniform magnetic field along the $x$ axis at the beginning 
of simulation. The field strength $B_0$ is specified by the parameter 
$\alpha$, the ratio of magnetic to thermal pressure, through 
\begin{equation}
B_0=3.22~\alpha^{1/2} \left({T\over 10 K}\right)^{1/2}
\left(\frac{n_{H_2,0}}{250\ {\rm cm}^{-3}}\right)^{1/2} (\mu~G). 
\label{fieldstrength}
\end{equation}
In units of the critical value $2\pi G^{1/2}$ (Nakano \& Nakamura
1978), the flux-to-mass ratio for the material inside the box is 
\begin{equation}
\Gamma_0 = 0.45 n_J^{-1} \alpha^{1/2}.
\label{centralgamma}
\end{equation}
Since we are interested in magnetic regulation of star formation,
we choose a relatively strong magnetic field with $\alpha=24$, 
corresponding to a mildly magnetically subcritical region with 
$\Gamma_0=1.1$ (or a dimensionless mass-to-flux ratio $\lambda
=0.91$) for $n_J=2$, which is close to the value of 1.2
used in the 2D simulations of LN04 and NL05. For the fiducial 
values for temperature and density, the initial field strength 
is $B_0=15.8\ \mu$G according to equation~(\ref{fieldstrength}); 
this value is not unreasonably high (see \S~\ref{taurus} for 
a discussion of the magnetic field in the Taurus region). We 
postpone an investigation of initially supercritical clouds to 
a future publication.  

Molecular clouds are highly turbulent, particularly for the 
relatively diffuse gas that we have assumed for the initial 
state. It is established that supersonic turbulence decays 
rapidly, with or without a strong magnetic field (e.g., Mac 
Low \& Klessen 2004). How the turbulence is maintained is 
uncertain (e.g., McKee \& Ostriker 2007). For the relatively 
small, parsec-scale region that is modeled here, a potential 
source is energy cascade from larger scales, which can 
in principle keep the turbulence in the sub-region for a 
time longer than the local crossing time. Ideally, this 
(and may other) form of driving should be included in a 
model, although it is unclear how this can be done in  
practice. We will take the limit that the turbulence decays 
freely, except for the feedback from protostellar outflows 
that are associated with forming stars. Other forms of 
driving will be considered elsewhere. Following the standard 
practice (e.g., Ostriker et al. 2001), at the beginning of 
the simulation ($t=0$) we stir the cloud with a turbulent 
velocity field of power spectrum $v_k^2 \propto k^{-3}$ in 
Fourier space.  

A useful time scale for cloud evolution is the global gravitational 
collapse time 
(Ostriker et al. 2001)
\begin{equation}
t_g= {L_J \over c_s} = 6.36 \times 10^6 \left(\frac{n_{H_2,0}} 
{250\ {\rm cm}^{-3}}\right)^{-1/2} ({\rm years}), 
\label{gravtime}
\end{equation}
which is longer than the free-fall time at the initial density, 
$t_{ff,0}=[3\pi/(32 G \rho_0)]^{1/2}$, by a factor of 3.27. In 
the presence of a strong magnetic field, as is the case in our
simulation, gravitational condensation is expected to occur 
preferentially along field lines, creating a flattened, sheet-like,
structure in which most star formation activities take place. 
Inside the sheet, the characteristic Jeans length is 
\begin{equation}
L_{s} \equiv {c_s^2 \over G \Sigma_0} = {1 \over \pi n_J} L_J 
=  0.195 \left(\frac{2}{n_J}\right) \left(\frac{T}{10 K}\right)^{1/2} 
\left(\frac{n_{H_2,0}}{250\ {\rm cm}^{-3}}\right)^{-1/2} {\rm pc}\  ,
\end{equation}
corresponding to a local gravitational collapse time 
\begin{equation}
t_{s} \equiv {L_{s} \over c_s} = {c_s \over G \Sigma_0} 
={1 \over \pi n_J} t_g \ = 1.01 \times 10^6 \left(\frac{2}{n_J}\right) 
\left(\frac{n_{H_2,0}} {250\ {\rm cm}^{-3}}\right)^{-1/2} ({\rm years}) ,
\end{equation}
where the column density $\Sigma_0$ is measured along the initial
magnetic field lines, $\Sigma_0 = \rho_0 L = n_J \rho_0 L_J$.
The sheet has a Jeans number $n_{s}\equiv L/L_{s} = \pi n_J^2$, 
corresponding to $12.6$ for $n_J=2$, which is close to the value 
of 10 adopted in the 2D simulations of LN04 and NL05. Note that 
the gravitational collapse time $t_s$ is longer than the free-fall 
time of the condensed sheet 
\begin{equation}
t_{\rm ff,s}=\left({3\pi \over 32 G \rho_s}\right)^{1/2},
\label{freefalltime}
\end{equation}
by a factor of 2.31, where the characteristic sheet density is
given by  
\begin{equation}
\rho_s = \frac{\pi G \Sigma_0^2}{2 c_s^2} = \frac{\pi^2 n_J^2}{2} 
\rho_0 \ .
\end{equation}

\subsection{Numerical Code}

We solve the equations that govern the cloud evolution using a 3D MHD 
code based on an upwind TVD scheme, with the ${\nabla\cdot {\bf B}}=0$ 
condition enforced through divergence cleaning after each time step. 
The base code is the same as the one  
used in the cluster formation simulations of Li \& Nakamura (2006) 
and Nakamura \& Li (2007), except that we have now included ambipolar 
diffusion in the induction equation. The ambipolar diffusion term is 
treated explicitly, which puts a stringent requirement on the time 
step (since it is proportional to the square of the grid size, Mac 
Low et al. 1995), especially in the lowest density regions. To 
alleviate the problem, we set a density threshold, $\rho_{AD}=0.1 
\rho_0$, below which the rate of ambipolar diffusion is reduced to 
zero. A similar approach was taken by Kudoh et at. (2007), and can 
be justified to some extent by increased ionization of low (column) 
density gas from UV background (McKee 1989). Even with this measure, 
it still takes more than a month to run a standard $128^3$ simulation 
for two global gravitational collapse time $t_g$ (or $4\pi$ times the 
sheet collapse time $t_s$) on the vector machine available to us. 
Since our focus is on the evolution of, and star formation in, the 
condensed gas sheet, to speed up the simulations, we follow the 
initial phase of cloud evolution on a coarser grid of $64^3$ up to 
a maximum density of $\sim 40~\rho_0$, before switching to the 
$128^3$ grid. To gauge the numerical diffusion of magnetic field on 
such a relatively coarse grid, we have followed the evolution of a 
model cloud in the ideal MHD limit (with the ambipolar diffusion 
term turned off; Model I0 in Table \ref{tab:model parameter}) up 
to $2 t_g (=12.6t_s)$, and found that the mass-to-flux ratio is 
conserved to within $1.5\%$, 
which seems to be much better than the two-dimensional 
higher resolution runs with ZEUS ($512^3$, Krasnopolsky \& Gammie 2005). 

A difficulty with grid-based numerical codes such as ours in 
simulating star formation in turbulent clouds is that, once 
the collapse of a piece of the cloud runs away to form the 
first star, the calculation grinds to a halt without special 
treatment. Because the bulk of the stellar mass is believed 
to be assembled over a rather brief (Class 0) phase of $\sim
3\times 10^4$~yrs (Andre et al. 1993) and the outflows are 
most active during this phase (much shorter than the fiducial 
characteristic time $t_s\sim 10^6$~yrs), we idealize the 
processes of individual star formation and outflow ejection 
as instantaneous, and treat them using the following simple 
prescriptions. When the density in a cell reaches a threshold 
value $\rho_{th}=10^3\rho_0$, we define around it a critical 
core using a version of CLUMPFIND algorithm 
(Williams, de Geus, \& Blitz 1994; 
see \S~\ref{core} for detail). The core identification 
routine is specified by a minimum threshold density $\rho_{\rm 
th, min}$ and the number of density levels $N$. They are set 
to $200~\rho_0$ and 5 respectively. The mean density of the 
core so identified is estimated at $\sim 10^5$~cm$^{-3}$, 
comparable to those of H$^{13}$CO$^+$ cores in Taurus studied 
by Onishi et al. (2002). We extract $30\%$ of the mass from 
the core (if its mass exceeds the local Jeans mass), and put 
it in a Lagrangian particle located at the cell center; its 
subsequent evolution is followed using the standard leap-frog 
method. The extracted mass fraction is 
motivated by the observations of J. Alves et al. (2007), 
who found a mass spectrum for dense cores that is similar to 
the stellar IMF in shape but more massive by a factor of $\sim 3$ 
(see also Ikeda, Sunada, \& Kitamura 2007). 
It is within the range of $25-75\%$ for the star formation
efficiency per dense core estimated theoretically by Matzner \& 
Mckee (2000). 

The material remaining in the core after mass extraction is 
assumed to be blown away in a two-component outflow: a bipolar 
jet of $30^\circ$ half-opening angle around the local magnetic 
field 
direction that carries a fraction $\eta$ of the total momentum, 
and a spherical component outside the jet that carries the 
remaining fraction. The total outflow momentum is assumed to 
be proportional to the stellar mass, with a proportionality 
coefficient of $P_*=35$~km/s. 
The prescriptions for star formation 
and outflow are similar to those used in Nakamura \& Li (2007). 
One difference is the use of CLUMPFIND here to define the critical 
core. Another is that, when an outflow reenters the computation 
box because of periodic boundary condition, we reduce its strength 
by a factor of ten if its speed exceeds $50~c_s$, to prevent 
it from strongly perturbing or even disrupting the condensed 
sheet. 

\section{Standard Model} 
\label{result}

We have run four models that include ambipolar diffusion, with 
parameters listed in Table \ref{tab:model parameter}. They
differ in random realization of the initial velocity field, 
relative strength of the jet and spherical outflow components,
and level of initial turbulence. An ideal MHD run was also 
performed for comparison. In this section, we focus on the 
model with initial turbulent Mach number ${\cal M}=3$ and 
jet momentum fraction $\eta=75\%$. It serves as the standard 
against which other models, to be discussed in the next 
section, are compared. 

The initial phase of cloud evolution is similar to those already 
presented in Krasnopolsky \& Gammie (2005), who carried out 
high-resolution, ideal MHD simulations of magnetically subcritical 
clouds in two dimensions, including self-gravity and an initially 
supersonic turbulence that decays freely, as in our simulation. 
The initial turbulence creates many shocklets that are oriented 
more or less perpendicular to the field lines; they appear as 
filaments of enhanced density and are the main sites of turbulence 
dissipation. As the turbulence decays, the fragments start to 
coalesce, eventually forming a single, approximate Spitzer sheet. 
In the ideal MHD limit, the sheet formation marks the end of 
cloud evolution. With ambipolar diffusion, it signals the beginning 
of a new phase---star formation\footnote{For a strong enough 
turbulence and/or initially inhomogeneous distribution of 
mass-to-flux ratio, stars can form before the formation of a 
well-defined sheet (see \S~\ref{discussion} for further
discussion).}. 
This latter phase is the focus of our discussion next.  

\subsection{Star Formation Efficiency}
\label{SFE}

\begin{figure}
\plotone{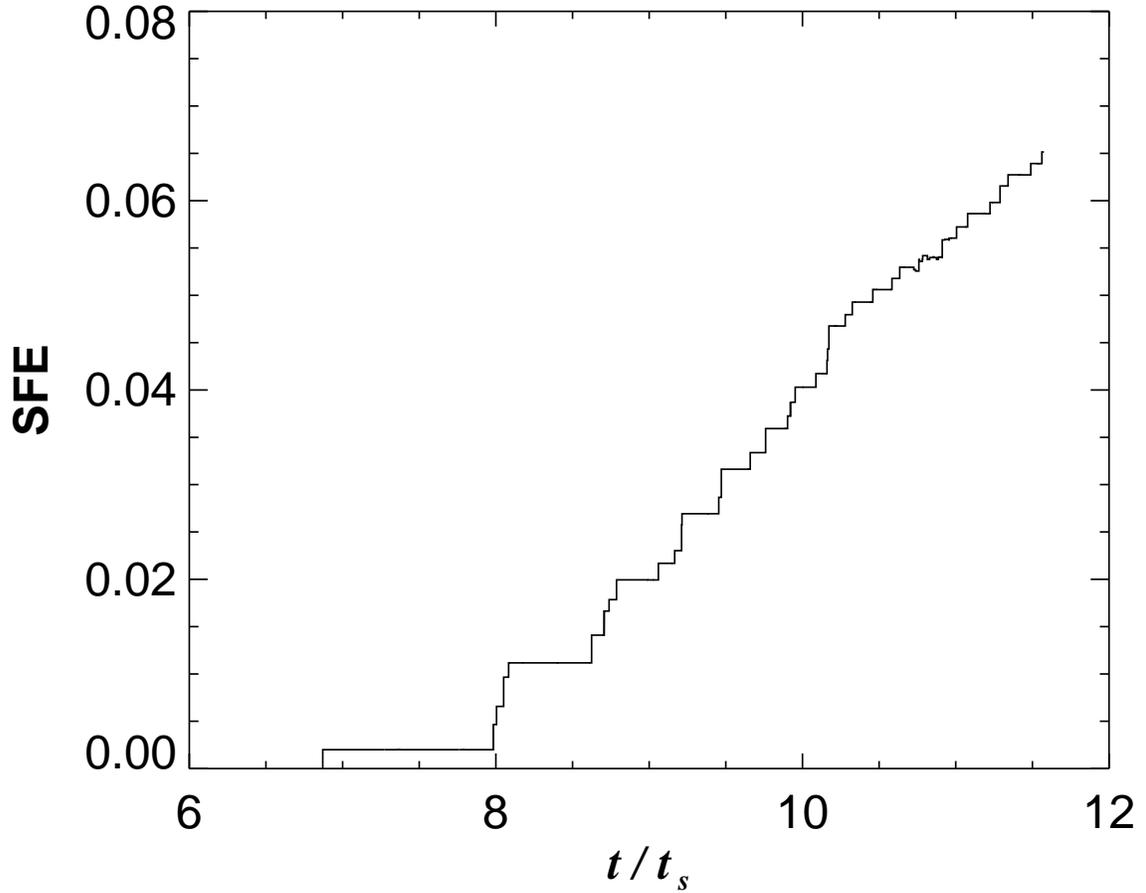}
\caption{Star formation efficiency (SFE) in the standard model as a 
function of time in units of the collapse time of the condensed 
sheet $t_s$. The slope of the curve indicates that less than 
$1~\%$ of the cloud mass is converted into stars per local 
free-fall time of the condensed sheet. }  
\label{fig:sfe}
\end{figure}

We begin the discussion with a global measure of star formation in the
standard model: the efficiency of star formation, defined as the ratio 
of the mass of all stars
to the total mass of stars and gas. The star formation efficiency (SFE 
hereafter) is plotted in Fig.~\ref{fig:sfe} as a function of time in 
units of the gravitational collapse time of the sheet $t_s$. Sheet 
formation 
is finished by the end of the first global collapse time $t_g=2\pi\ 
t_s$. Soon after, the first star appears, around $t \approx 7~t_s$. 
By the end of simulation at $t=11.4~t_{s}$ (nearly $2~t_g$), 37 stars 
have formed, yielding an accumulative SFE of 6.5\%. The average mass 
of a star is close to $0.45 M_\odot \left(T/10 K\right)^{3/2}\left(n_{H_2,0}
/250 {\rm cm}^{-3} \right)^{-1/2}$, typical of low-mass stars. The 
rate of star formation is given by the slope of the SFE curve, which 
is estimated at $\sim 1.5\%$ per $t_s$. At this 
rate, it would take $\sim 65~t_s$ to deplete the gas due to star 
formation. The gas depletion time corresponds to $\sim 10~t_g$ or 
$\sim 33$ times the free-fall time $t_{{\rm ff},0}$ at the {\it 
initial} (uniform) density $\rho_0$. In other words, 
the star formation efficiency per {\it initial} free fall time 
(Krumholz \& McKee 2005) is SFE$_{{\rm ff},0}\sim 3\%$. Interestingly,
this rate is comparable to that of cluster formation in 
protostellar outflow-driven turbulence (Nakamura \& Li 2007). 
However, a large fraction of the gas is at a much higher density
than the initial density $\rho_0$ after the formation of the 
condensed sheet. It is more appropriate to measure the star 
formation rate using the free fall time at the characteristic 
sheet density $\rho_s=2 \pi^2\ \rho_0$, which is $t_{{\rm ff},s}= 
0.43\ t_s$ (see equation~[\ref{freefalltime}]), yielding 
SFE$_{{\rm ff},s}\sim 0.7\%$. To help understand the rather slow 
and inefficient star formation, we next examine in some detail 
the cloud structure and dynamics at $t=11~t_s$, representative of 
the post-sheet formation phase of cloud evolution when stars 
are forming steadily, starting with overall mass distribution 
as revealed by column density. 

\subsection{Column Density Distribution}

\begin{figure}
\plotone{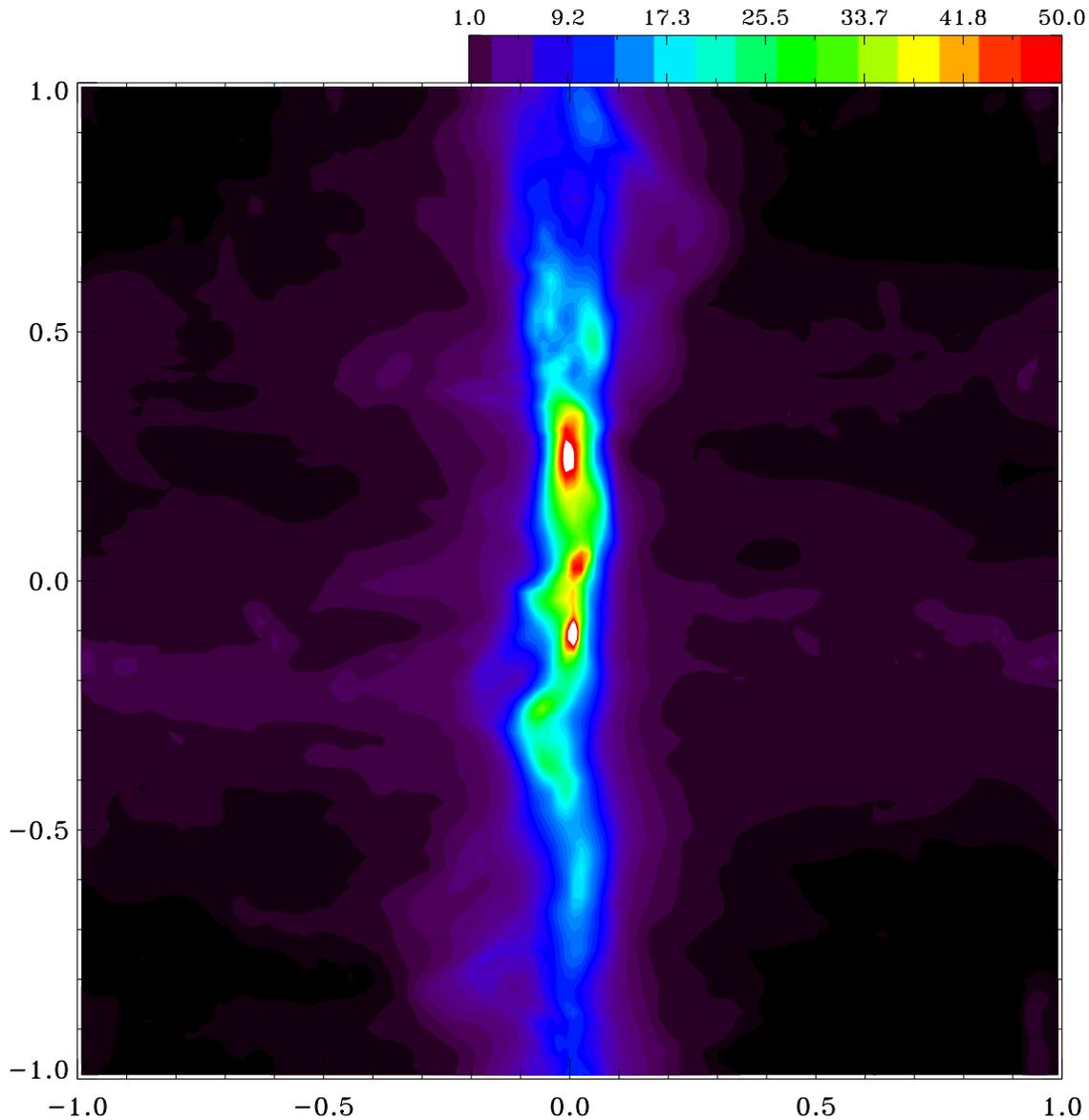}
\caption{An edge-on view of the condensed sheet at a representative
time $t=11~t_s$, showing the column
density distribution along the $y-$axis, which is perpendicular to
the direction of initial magnetic field. The condensation is surrounded
by a diffuse halo, with some spurs streaking away perpendicular to
the sheet. The length is in units of $L_J$ (the Jeans length of
the initial uniform state), and column density in the color-bar
in units of $\rho_0 L_J$. The three lowest contours have values of 
0.05, 0.15, and 0.5, respectively, and are not shown in the
color-bar. They are plotted to highlight the structure of the
diffuse gas.} 
\label{colden_y}
\end{figure}

\begin{figure}
\plotone{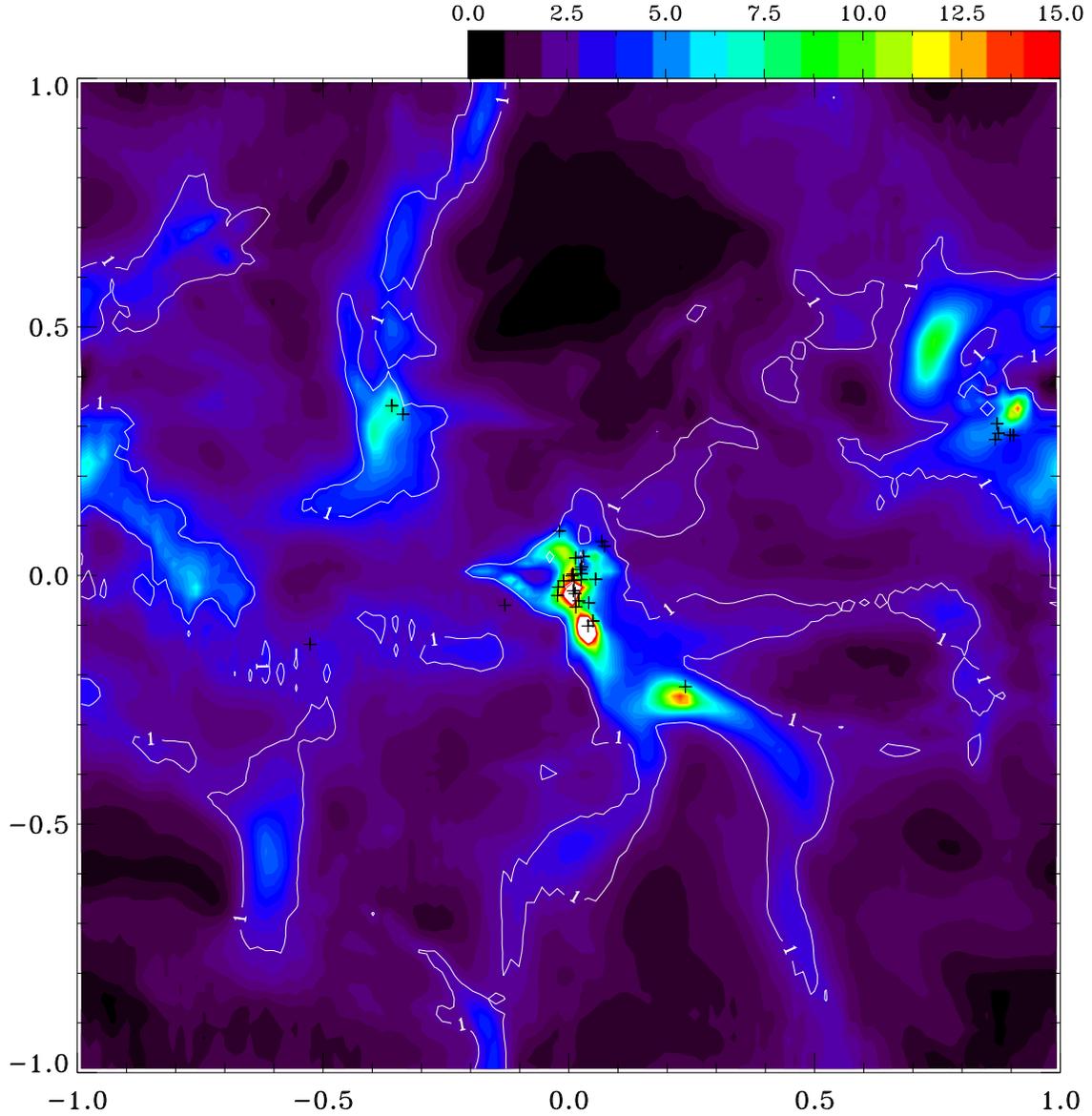}
\caption{Face-on view of the same condensed sheet as in
  Fig.~\ref{colden_y}, showing the column
density distribution along the $x-$axis, the direction of initial 
magnetic field. Superposed are stellar positions (crosses) and 
contours of critical average mass-to-flux ratio (${\bar \lambda}
=1$, see equation~[\ref{AverageLambda}]). The contours divide the 
magnetically supercritical material capable of star formation from 
the ambient subcritical material sterile to star formation.}  
\label{colden_critical}
\end{figure}

The cloud material is expected to condense along the magnetic field
lines into a flattened structure. This expectation is borne out by
Figs.~\ref{colden_y} and \ref{colden_critical}, which show the 
distributions of column density along, respectively, the $y$- and 
$x$-axis (the direction of the initial magnetic field). Viewed
edge-on, the structure resembles a knotty, wiggly filament 
(Fig.~\ref{colden_y}). The dense filament is sandwiched by a 
low-density ``halo,'' which contains a number of spurs more or
less perpendicular to the filament. The spurs are roughly aligned 
with the global magnetic field. They resemble the diffuse, fuzzy 
$^{12}$CO streaks that are frequently observed to stream away 
from dense filaments (Goldsmith et al. 2008). Similar spurs are 
seen in the recent SPH-MHD simulations of Price \& Bate (2008). 

The face-on view of the condensed structure is quite different 
(Fig.~\ref{colden_critical}). It appears highly fragmented, with 
low-density voids as well as high-density peaks. Some 
over-dense regions cluster together to form a short filament; 
others are more scattered. The actual appearance of the cloud 
to outside observers will depend, of course, on the viewing 
angle. It can be filament-like (as in Fig.~\ref{colden_y}) or, 
more likely, patchy (as in Fig.~\ref{colden_critical}). Superposed 
on the face-on column density map is the distribution of the 
33 stars that have formed up to the time plotted. With few 
exceptions, they follow the dense gas closely. Most of the stars
are found in small groups, which is broadly consistent with the 
distribution of young stars in Taurus (Gomez et al. 1993). Since 
the clouds is initially 
magnetically subcritical everywhere, in order for any localized 
region to collapse and form stars, it must become supercritical 
first. 
%
% plots recentered, and in units of Lg rather than Ls
%

\begin{figure}
\plotone{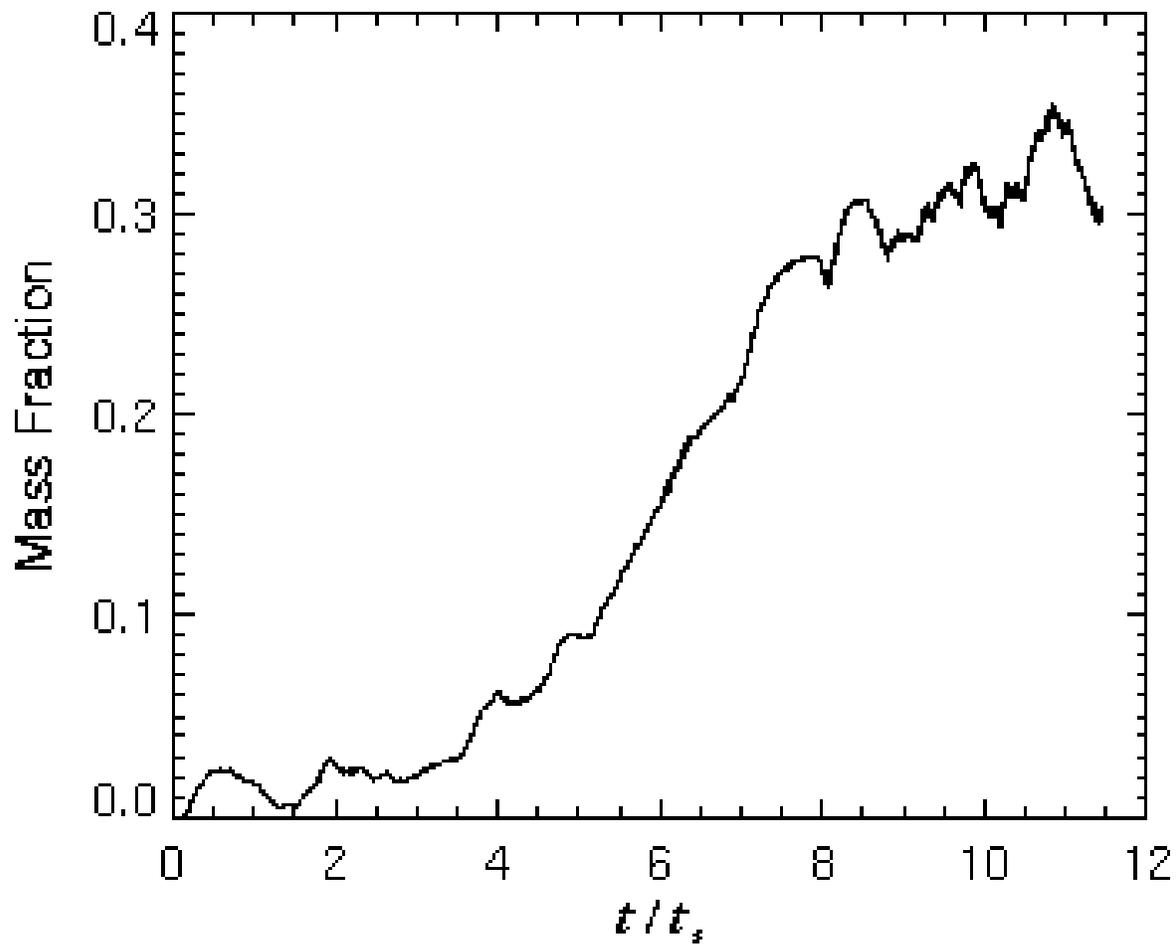}
\caption{Mass fraction of magnetically supercritical material
created in the initially subcritical cloud of the standard model
through ambipolar diffusion. The mass fraction of supercritical 
gas increases gradually initially, reaching a plateau at late 
times. 
}  
\label{fig:supercritical}
\end{figure}

A rough measure of magnetic criticality is provided by the 
average mass-to-flux ratio $\bar{\lambda}$ normalized to
the critical value,
\begin{equation}
\bar{\lambda}(y,z) = \int_{-L/2}^{+L/2} 2\pi G^{1/2}
\rho/B_x dx.  
\label{AverageLambda}
\end{equation}
Contours of $\bar{\lambda}=1$ are plotted in Fig.~\ref{colden_critical}, 
showing that most of the dense regions are indeed magnetically 
supercritical, and thus capable of forming stars in principle. 
The supercritical regions enclosed within the contours contain 
a significant fraction ($34\%$) of the total mass. 
The mass fraction of supercritical gas increases gradually with 
time initially, and remains more or less constant at later times 
($t\gtrsim 6~t_s$), as shown in Fig. \ref{fig:supercritical}.
This value provides an upper limit to the efficiency of 
star formation.
If the supercritical regions collapse to form stars within a local 
free fall time, the star formation rate would be much higher 
than the actual value. An 
implication is that the bulk of the supercritical material 
remains magnetically supported to a large extent and is rather 
long lived. Only its densest parts (i.e., dense cores), 
which contain a small fraction of the supercritical mass, 
directly participate in star formation at any given time. 
The longevity is characteristic of the (mildly) supercritical 
regions that are produced through turbulence-accelerated 
ambipolar diffusion in nearly magnetically critical clouds. 
It is in contrast with the highly supercritical material 
in weakly magnetized clouds, which is expected to collapse 
promptly unless supported by turbulence. This same behavior 
was found in 2D calculations (NL05).

To quantify the mass distribution further, we plot in Fig.~\ref{colden_PDF} 
the probability distribution function (PDF) of the column density 
along the $x-$axis, in the direction of initial magnetic field. The 
PDF peaks around $\Sigma \sim 1.7~\rho_0 L_J$ (where $\rho_0$ and 
$L_J$ are the density and Jeans length of the initial, uniform state), 
close to the average value of $2~\rho_0 L_J$. The distribution can be 
described reasonably well by a lognormal distribution, although there 
is significant deviation at both low and high density end. The 
considerable spread in column density may be surprising at the 
first sight, since one may expect little structure to develop in 
a subcritical sheet dominated by a strong, more or less uniform 
magnetic field, especially since the initial mass distribution is 
also chosen to be uniform. This expectation is consistent with the 
column density PDF of the ideal MHD counterpart of the standard 
model (plotted in the same figure for comparison), which is much 
narrower. The key difference is that ambipolar diffusion turns an 
increasingly large fraction of the initially subcritical cloud into 
supercritical material (until saturation sets in at late times, 
see Fig.~\ref{fig:supercritical}). Once produced, the supercritical 
material behaves drastically differently from the subcritical 
background. It is easier to compress by an external flow because 
its magnetic field is weaker. The compressed supercritical region
also rebounds less strongly, because the magnetic forces driving 
the rebound are cancelled to a larger extent by the self-gravity 
of the region. More importantly, the supercritical regions can 
further condense without external compression, via self-gravity. 
The accumulation of supercritical material and subsequent  
(magnetically-diluted) gravitational fragmentation in such material 
are responsible for the broad column density PDF. 
These key aspects of the problem are {\it not} captured by the 
ideal MHD simulation, but are well illustrated in 2D simulations 
that include ambipolar diffusion (e.g., NL05; Kudoh \& Basu 
2006). Other aspects are better discussed in 3D, however. These 
include the dependence of various physical quantities on volume 
density, to which we turn our attention next. 

\subsection{Density Stratification} 

Our model cloud is strongly stratified in density. It contains dense 
cores embedded in a condensed sheet which, in turn, is surrounded 
by a more diffuse halo. The relative amount of mass distributed at 
different densities is displayed in Fig.~\ref{lognorm_density},
along with that for the ideal MHD case and for a Spitzer sheet. 
Whereas the mass distribution for the ideal MHD case resembles 
that of the Spitzer sheet closely, the same is not true for 
the standard model with ambipolar diffusion: its mass distribution 
is considerably broader, with more material at both the low and 
high density end. The former indicates that the diffuse material 
outside the sheet is dynamically active, as we show below. The 
excess at the high density end is produced, on the other hand, 
primarily by ambipolar diffusion, which enables pockets of 
the sheet material to condense across magnetic field lines 
gravitationally. The bulk ($62.5\%$) of the cloud material 
resides in the condensed sheet, which we define somewhat 
arbitrarily as the material in the density range between 
$5~\rho_0$ and $50~\rho_0$, corresponding to a range in 
H$_2$ number density between $1.25\times 10^3$~cm$^{-3}$ and 
$1.25\times 10^4$~cm$^{-3}$ (assuming $n_{H_2,0}=250$~cm$^{-3}$). 
The remaining cloud material is shared between the diffuse halo 
($30.0\%$), defined 
loosely as the material with density below $5\ \rho_0$, and 
dense cores ($7.5\%$), defined as the material denser than
$50 \rho_0$ (see \S~\ref{core} below). 

Our model cloud is also strongly stratified in turbulent speed. 
Fig.~\ref{rms_vel} shows the mass weighted rms velocity as a 
function of density. At densities corresponding to the 
condensed sheet ($5-50~\rho_0$), the turbulent Mach number is 
about 2. The moderately supersonic turbulence persists in the 
condensed sheet for many local free fall times, despite the 
absence of external driving. It is driven by a combination 
of outflows (through either direct impact or MHD waves) and 
ambipolar diffusion-induced gravitational acceleration. 
The modest increase in rms velocity toward the high density 
end is because the denser gas is typically closer to star 
forming sites and is thus more strongly perturbed by outflows. 
The much higher level of turbulence in the diffuse halo is 
more intriguing. One may expect little turbulence to remain at 
the representative time $t=11~t_s$ 
shown in the figure, which is well beyond the time ($t\sim 6\ t_s$) 
of sheet formation --- a product of turbulence dissipation. The 
expectation is consistent with the ideal 
MHD simulation, which shows that, in the absence of external
driving, the turbulent motions die down quickly everywhere. 
The turbulence decay is changed by ambipolar diffusion 
which, accelerated by the initial strong turbulence, enables 
stars to form soon after sheet formation. The stars stir 
up the halo via protostellar outflows in at least two ways. 
First, the outflows are powerful enough to disrupt the 
star-forming cores, break out from the condensed sheet, and 
inject energy and momentum directly into the halo. Second, a 
fraction of the outflow energy and momentum is also 
deposited in the sheet, where gas motions can shake the 
field lines and send into the halo MHD waves whose amplitudes 
grow as the density decreases (Kudoh \& Basu 2003); the waves 
help maintain the turbulence in the diffuse region, particularly 
in the cross-field direction.

Even though the turbulence speed is relatively high in the halo, 
it is still sub-Alfv\'enic. This is shown in 
Fig.~\ref{ave_MA}, where the average Alfv\'en Mach 
number $M_A$ is plotted as a function of density. The flow 
is moderately sub-Alfv\'enic at densities between $\sim 0.3$ 
and $\sim 10~\rho_0$, with $M_A$ within $20\%$ of $0.5$. The
Alfv\'en Mach number 
drops slowly towards lower densities, indicating that the 
lower-density gas becomes increasingly more magnetically
dominated in our model cloud. A caveat is that, in real 
clouds such as the Taurus complex, the dynamics of the 
low-density halo on the parsec-scale of our modeled region 
are expected to be tied to that of the larger complex. It 
is likely for the diffuse gas to receive energy cascaded 
from larger scales and become more turbulent than suggested 
by Fig.~\ref{ave_MA}. Some of this energy may even be fed 
through Alfv\'en waves into the condensed sheet and increase 
its velocity dispersion as well. The increase is not expected 
to be large, however, since the Alfv\'en speed becomes 
transonic at high densities (see Fig.~\ref{ave_MA}), where 
highly supersonic motions should damp out quickly through 
shock formation. In future refinements, it will be 
desirable to include external driving of the halo material, 
which may modify the way stars form, especially at early 
times (see discussion in \S~\ref{discussion}). Nevertheless, 
we believe that the decrease of turbulent speed with increasing 
density found in our simulation (and earlier in Kudoh \& 
Basu 2003) is a generic feature of the magnetically 
regulated star formation in three dimensions. It is broadly 
consistent with the observed trend (e.g., Myers 1995). 

Another generic feature is the distribution of magnetic field 
strength as a function of density. The distribution is shown 
in Fig.~\ref{ave_B}. There is a clear trend for the average 
field strength to remain more or less constant at relatively 
low densities, before increasing gradually towards the high 
density end. A similar trend is inferred 
in Zeeman measurements of field strength over a wide range of 
densities, and has been interpreted to mean that the initial 
stage of cloud condensation leading to star formation is 
primarily along the field lines, which increases the density 
without increasing the field strength (Heiles \& Crutcher 
2005). This is precisely what happens in our model, 
where the initial accumulation of diffuse material into the 
condensed sheet is mostly along the field lines, and further 
condensation inside the sheet is primarily in the cross-field 
direction. The cross-field condensation leads to an increase 
in field strength with density, although by a factor that is 
less than that expected from flux-freezing because of ambipolar 
diffusion. 

\subsection{Sheet Fragmentation: Turbulent or Gravitational?} 

The star formation activities in our model are confined mostly to 
the condensed sheet. Its dynamics hold the key to understanding the 
magnetically regulated star formation. An overall impression of
the sheet can be gained from Figs.~\ref{colden_y} and 
\ref{colden_critical},
which showed an edge-on and face-on view of the sheet in column 
density. The sheet has a highly inhomogeneous 
mass distribution which is, of course, a pre-requisite for active
star formation. To illustrate the clumpy mass distribution further, 
we plot in Fig.~\ref{den_pot_vel} the distribution of the density  
in the ``mid-plane'' of the sheet, defined as the $yz$ plane 
(perpendicular to the initial magnetic field along the $x$ axis) 
that passes through the minimum of gravitational potential. The 
density distribution spans seven orders of magnitude, from 
$10^{-4}~\rho_0$ (the density floor adopted in the computation) to 
$10^3~\rho_0$ (the threshold for Lagrangian particle generation).
The question is: how is the density inhomogeneity created and
maintained? 

A widely discussed possibility is turbulent fragmentation, where 
regions of high densities are created by converging flows (e.g., 
Padoan \& Nordlund 2002; Mac Low \& Klessen 2004). There is indeed 
a random velocity field in the mid-plane, as shown by arrows in 
the figure. However, the flow speed is typically small, comparable 
to the sound speed in most places. 
It is unlikely for such a weak turbulence to create the high density 
contrast shown in Fig.~\ref{den_pot_vel} by itself. This conclusion 
is strengthened by the presence of a strong magnetic field, which 
cushions the flow compression. Indeed, the turbulent speeds in the 
plane of the sheet are significantly below the approximate
magnetosonic speed $v_{ms}=[B_x^2/(4\pi\rho)+c_s^2]^{1/2}$ in
most places, and are thus not expected to create large density 
fluctuations; they are essentially a collection of magnetosonic 
waves. The low level of turbulence points to gravity rather than 
flow motions as the primary driver of fragmentation in the sheet. 
%
% (1) potential loop holes... what if large compression in individual 
% regions involving a small fraction of mass??? i.e., tail effect???
% 
% Indeed the case at freshly shocked sites by active outflows
%

Gravity is expected to be dynamically important in the sheet, a structure 
formed by gravitational condensation along field lines in the 
first place. The fact that the sheet has a dimension much larger than
the thickness indicates that it contains many thermal Jeans masses. 
If the thermal pressure gradient was the main force that opposes 
the gravity, one would expect the bulk of the sheet to collapse 
in one gravitational time $t_s$. And yet, the vast majority of 
the sheet material remains uncollapsed after at least $\sim 5~t_s$. 
The longevity indicates that the 
gravitational fragmentation proceeds at a much reduced 
rate. The reduction is due, of course, to magnetic forces, 
which cancel out the gravitational forces in the plane of 
the sheet to a large extent (Shu \& Li 1997, Nakamura et al. 
1999). The near cancellation 
can be inferred from Fig.~\ref{den_pot_vel}, where contours of
gravitational potential are plotted along with the density
and velocity vectors. If the gravitational forces were 
unopposed, one would expect the material to slide down the 
potential well, and picking up speed towards the bottom, 
with the infall speed reaching highly supersonic values. 
For example, for a potential drop of 8~$c_s^2$ (say between 
contours labeled  ``$-8$'' and ``$-16$''), an increase in speed 
by 4~$c_s$ is expected. These expectations are not met. The 
actual velocity field appears quite random, and generally 
transonic. The slow random motions in a rather deep 
gravitational potential well provide a clear indication 
for magnetic regulation of the sheet dynamics. Active 
outflows also play an important role in generating the density 
inhomogeneities observed in the condensed sheet. Their effects 
are discussed in \S~\ref{outflow} below. 

\subsection{Comparison to 2D Calculations}
\label{comparison}

There are a number of differences between the 3D simulation and the 
previous 2D calculations. The 2D treatment does not account for the 
mass stratification along the field lines, and thus cannot 
address the important issue of dynamical coupling between the 
condensed material and the diffuse halo. A related difference is
that, in 3D, there is an initial phase of mass settling to form the 
sheet, which is is absent in 2D. Furthermore, the mass of each star 
in the 2D models is held fixed (typically at $0.5~M_\odot$); in 3D, 
it is fixed at $30\%$ of the mass of the criticl core that forms 
the star. In addition, the prescription for outflow
\footnote{We note that the spherical component of the outflow 
in the standard 3D model carries a momentum close to  
(within $15\%$ of)  that of the 2D outflow in the plane of 
mass distribution in the standard model of NL05.}
 and the degree 
of initial cloud magnetization are also somewhat different. Despite 
these differences, the results of the 2D and 3D models are in 
remarkably good agreement. 

The most important qualitative agreement is that, once settled 
along the magnetic field lines, the condensed sheet material 
produces stars at a slow rate. In the standard 2D model of NL05, 
about $5.3\%$ of the total mass is converted into stars within 
a time interval of $3.35~t_s$ (see their Fig.~1), corresponding 
to a star formation rate of $\sim 1.6\%$ per sheet gravitational 
collapse time $t_s$. This rate is nearly identical to the value 
of $\sim 1.5\%$ estimated in \S~\ref{SFE} for the standard 3D 
model (which is also close to that of the 3D ${\cal M}=10$ model, 
see \S~\ref{other} below). The (face-on) mass distributions 
are also similar in 2D and 3D models. In both cases, dense, 
star-forming cores are embedded in moderately supercritical 
filaments which, in turn, are surrounded by more diffuse, 
magnetically subcritical material (compare 
Fig.~\ref{colden_critical} with the lower panels of 
Fig.~2 of NL05). More quantitatively, the mass fraction of 
supercritical material increases gradually initially and 
reaches a plateau of $\sim 30\%$ in both 2D and 3D models. The 
similarity extends to the turbulent speed of the condensed 
material, which is about twice the sound speed in both cases 
at late times (see Fig.~\ref{rms_vel} and Fig.~11 of NL05). 

The similarities indicate that 2D calculations capture the 
essence of the slow, magnetically regulated star formation 
in moderately subcritical clouds. The same conclusion was
drawn by Kudoh et al. (2007) based on 3D simulations of 
weakly perturbed magnetically subcritical sheets. The 
2D calculations have the advantage of being much faster to 
perform. They can be done with higher spatial resolution 
($512^2$ as compared to the $128^3$ used in this paper) 
and in larger numbers. The number of 3D simulations is 
necessarily more limited, especially because of the 
small time step required by our treatment of ambipolar 
diffusion using an explicit method. Nevertheless, we are 
able to perform several additional simulations, to check 
the robustness of the standard model, and to help elucidate 
the effects of two key ingredients: initial turbulence and 
outflow feedback.  

\section{Other Models} 
\label{other}

\subsection{Initial Turbulence}
\label{turbulence}

Even though the initial turbulence dissipates quickly, it seeds 
inhomogeneities in mass distribution and mass-to-flux ratio 
that are expected to affect star formation later on. To illustrate 
the effects of initial turbulence, we focus on a model identical 
to the standard model, except for a higher initial turbulent Mach 
number ${\cal M}=10$ (instead of $3$). The overall pattern of 
cloud evolution follows that seen 
in the standard model: dissipation of turbulence leads to 
gravitational settling of cloud material onto a condensed sheet, 
the densest parts of which collapse to form stars; the stars, 
in turn, stir up the sheet and its surrounding halo through 
protostellar outflows. One difference is that the stronger 
turbulence initiates star formation at an earlier time, as seen 
in Fig.~\ref{sfetwo}. The first star forms around $\sim 4.5~t_s$ 
in the ${\cal M}=10$ case, compared to $\sim 6.8~t_s$ for 
${\cal M}=3$. Nevertheless, their rates of star formation are 
comparable, with values between $\sim 0.5\%$ and $\sim 1.0\%$ 
per free fall 
time $t_{ff,s}=0.43\ t_s$ in both cases. The earlier start, but 
still low rate, of star formation in the more turbulent case 
is consistent with the turbulence-accelerated, magnetically 
regulated scenario that we advocate based on previous 2D 
calculations (LN04, NL05; see also recent 3D simulations of 
Kudoh \& Basu 2008). 

The basic structure and kinematics of the initially more turbulent 
cloud are also similar to those of the standard model cloud. In 
both cases, the condensed sheet is highly fragmented but dynamically 
``cold,'' with moderately supersonic random motions. It is surrounded 
by a halo moving at highly supersonic, but sub-Alfv\'enic, speeds. 
An interesting difference is that the sheet in the ${\cal M}=10$ 
case appears thicker, as illustrated in Fig.~\ref{colden_y_M10},
where an edge-on view of the column density distribution (along 
the $y$-axis, perpendicular to the initial magnetic field 
direction) is plotted for a representative time $t=8~t_s$. It is 
to be compared with Fig.~\ref{colden_y} for the standard model. 
The main reason for the larger apparent thickness is that the 
condensed sheet, while still thin intrinsically, is more warped. 
An example of warps is shown in Fig.~\ref{den_Bfield_M10}, where 
the density distribution in a plane perpendicular to the $y$-axis 
is plotted; superposition of such warps in different parts of 
the sheet along the line of sight gives rise to the more 
puffed-up appearance.

Associated with the sharp warp of condensed material in 
Fig.~\ref{den_Bfield_M10} is a strong 
distortion of magnetic field lines. Both are likely driven by 
protostellar outflows, which are the main source of energy 
that drives the cloud evolution at late times. In particular, 
outflows ejected at an angle to the large-scale magnetic field 
can excite large-amplitude Alfv\'en waves, which perturb the 
field lines in the halo that thread other parts of the condensed 
sheet. The large Alfv\'en speed in the diffuse halo allows 
different parts of the sheet to communicate with one another 
quickly, as emphasized recently by Shu et al. (2007) in the 
context of poloidally magnetized accretion disks. The 
communication raises the interesting possibility of global, 
magnetically mediated, oscillations for the condensed material 
that may have observational implications. This issue warrants 
a closer examination in the future.  

We conclude by stressing that condensations produced by gravitational
settling along field lines do not have to appear highly elongated, 
even when viewed edge-on. They can be further distorted by outflows 
from stars formed in neighboring condensations or other external 
perturbations that are not included in our current calculations. We 
have experimented with simulations that allow active outflows to 
re-enter the simulation box, and found that the condensed sheet 
became more puffed up and, sometimes, even completely disrupted. 

\subsection{Outflow Feedback}
\label{outflow}

Protostellar outflow plays an important role in our model. It 
limits the amount of core mass that goes into a forming star 
and injects energy and momentum back to the cloud. In our 
standard model, we have assumed that $75\%$ of the outflow 
momentum is carried away in a bipolar jet (of $30^\circ$ in
half opening angle), and the 
remaining $25\%$ in a spherical component. Here, we contrast 
the standard model with one that has a stronger spherical 
component (carrying $75\%$ of outflow momentum). The star
formation efficiencies (SFEs) of both models are plotted 
in Fig.~\ref{sfedependm3}. It is clear that the stronger 
spherical component of outflow has reduced the rate (and
efficiency) of star formation significantly, by a factor of 
$\sim 2.3$. This trend is the opposite of what we found in 
the case of cluster formation in dense clumps that are
globally magnetically supercritical (Nakamura \& Li 2007):  
for a given total outflow momentum, more spherical outflows 
tend to drive turbulence on a smaller scale that dissipates 
more quickly, leading to a higher rate of star formation. 

There are a couple of reasons for the seemingly contradictory 
result. First, whereas the supercritical cluster-forming 
clumps are supported mainly by turbulent motions, the bulk 
of the star-forming condensations in subcritical or nearly 
critical clouds are magnetically supported (in the cross-field 
direction). As a result, faster dissipation of 
turbulence leads to more rapid star formation in the former, 
but not necessarily in the latter. Second, the jet component, 
while crucial for supporting the more massive, rounder 
cluster-forming clump, has a more limited effect on the flattened 
condensation: it escapes easily from the sheet (and quickly 
out of the simulation box) and its energy is mostly ``wasted'' 
as far as driving turbulence in the star-forming, condensed 
material is concerned. The spherical component, in contrast, 
impacts the condensed material more directly, especially in 
the vicinity of a newly formed star. The dispersal of the 
residual core material after star 
formation to a wider region appears to be the main reason 
for the lower star formation rate in the case of stronger 
spherical component: it takes longer for the more widely 
dispersed material to recondense and form stars. The stars 
in the stronger spherical component model are less clustered 
than those in the standard model, consistent with the above 
interpretation.  

\section{Dense Cores}
\label{core}

\subsection{Identification}

A variant of the CLUMPFIND algorithm of Williams et al. (1994) is
ussed for core identification. We apply their ``friends-of-friends'' 
algorithm to the density data cube in regions above a threshold 
density, $\rho_{\rm th, min}$, and divide the density distribution 
between $\rho_{\rm th, min}$ and $\rho_{\rm th, max}$ into 15 bins 
equally spaced logarithmically. The maximum is set to $\rho_{\rm th, 
max}=10^3\ \rho_0$, the density above which a star (Lagrangian 
particle) is created if the core mass exceeds the local Jeans 
mass. The minimum is set to $\rho_{\rm th, min}=50\ \rho_0$, 
corresponding to $n_{H_2}=1.25 \times 10^4$~cm$^{-3}$ for the 
fiducial value of initial density. We include only those 
``spatially-resolved'' cores containing more than 30 cells in 
the analysis, to ensure that their properties are determined with 
reasonable accuracy. The minimum core mass is thus $M_{c, \rm min}
\equiv \rho_{\rm th, min} \times 30\Delta x\Delta y\Delta z=7.19 
\times 10^{-2} M_{J,s}$, where $M_{J,s}\equiv \rho_s L_s^3 
\simeq2.5M_\odot\left(T/10 K\right)^{3/2}\left(n_{H_2,0}/250 
{\rm cm}^{-3} \right)^{-1/2}$ is the Jeans mass of the Spitzer 
sheet. 

In our standard simulation, the cloud material settles into a large 
scale gas sheet by the time $t\sim 6\ t_s$. Soon after, a dense core
collapses in a run-away fashion to form the first star, signaling 
the beginning of active star formation. We will limit our analysis 
of dense cores to this star-forming phase, which complements the 
analyses of, e.g., Gammie et al. (2003) and Tilley \& Pudritz (2007), 
which focus on the phase prior to the runaway collapse. 

There are typically a handful of dense cores in our standard model at 
any given time (see Fig.~\ref{colden_critical}), which is too small 
for a statistical analysis. To enlarge the core sample, we rerun the 
standard model with a different realization of the initial turbulent 
velocity field, and include all ``resolved'' cores at six selected 
times $t=6.0$, 7.0, 8.0, 9.0, 10.0, and 11.0~$t_s$ from both runs. 
The total number of cores is 72. Since the majority of them do not 
survive for more than 1~$t_s$ (they do last for several local
free-fall times typically), there is relatively little overlap in 
dense cores from one time to another. We treat all cores independently 
regardless whether they appear at more than one selected time or 
not. They have densities in the range 10$^4\sim 10^5$ cm$^{-3}$ 
and radii of $0.04 \sim 0.12$ pc, corresponding roughly to 
H$^{13}$CO$^+$ or N$_2$H$^+$ cores.

\subsection{Core Shape}

We compute the shape of a core using the eigenvalues of the 
moment-of-inertia tensor:
\begin{equation} 
I_{ij} \equiv \int \rho x_i x_j dV  \ ,
\end{equation}
where the components of position vector, $x_i$ and $x_j$, 
are measured relative to the center of mass.
We compute the lengths of the three principal axes, $a, b,$ and $c$
($a\ge b \ge c$) by dividing the eigenvalues of the tensor, 
$M_ca^2$, $M_cb^2$, and $M_c c^2$, by the core mass $M_c$.
The core shape is then characterized by the axis ratios 
$\xi\equiv b/a$ and $\eta \equiv c/a$.

Figure \ref{fig:coreshapenew} shows the distribution of axis ratios 
of the dense cores in our sample. It is clear that the cores
are generally triaxial; they show little clustering around either the 
prolate ($b/a=c/a<1$) or oblate ($b/a=1, c<1$) line in the figure. 
This result is perhaps to be expected, since the cores are distorted 
by external turbulent flows and occasionally outflows, neither of 
which are symmetric. 
Following Gammie et al. (2003), we divide the $\xi$-$\eta$ plane 
into three parts: a prolate group which lies above the line
connecting $(\xi, \eta)=(1.0, 1.0)$ and (0.33, 0.0), an oblate group
which lies below the line connecting $(\xi, \eta)=(1.0,1.0)$ and 
(0.67, 0.0), and a triaxial group which is everything else.
Of the 72 identified cores, 43\% (31) are prolate, 44\% (32) are
triaxial, and 13\% (9) are oblate.  These fractions are in general 
agreement with Gammie et al. (2003). One  
difference is that the ratio $\eta=c/a$ tends to be smaller for 
our cores [see Figure 10 of Gammie et al. (2003)]. It is likely 
because dense cores in our model are embedded in a condensed sheet, 
which has already settled 
gravitationally along the field lines, although part of the 
difference may also come from the different methods used in
core identification. Note that more massive cores tend to be more 
flattened along their shortest axis. This trend lends support to 
the idea that gravitational squeezing plays a role in core 
flattening. 

Another trend is that more massive cores tend to be more oblate. 
For example, 9 out of 15 (60\%) cores having masses more than 1.5 
times the average core mass 
($\left<M_c\right>=0.93~M_{J,s}$) 
lie 
below the line connecting $(\xi, \eta)=(1.0,1.0)$ and (0.5, 0.0)
(the dotted line in Fig.~\ref{fig:coreshapenew}). Less massive cores
tend to be more prolate. Their shapes may be more susceptible to
external influences, such as ambient flows. 

\subsection{Virial Analysis}

Virial theorem is useful for analyzing the dynamical states of dense 
cores. We follow Tilley \& Pudritz (2007) and adopt the virial 
equation in Eulerian coordinates (McKee \& Zweibel 1992): 
\begin{equation}
{1 \over 2} \ddot{I} + {1 \over 2} \int _S \rho r^2 
\mbox{\boldmath $v$} \cdot d\mbox{\boldmath $S$} 
= U+K+W+S+M+F \ ,
% = {1 \over 2} \ddot{I'}
\label{eq:virial}
\end{equation}
where
$$
%\begin{eqnarray}
I=\int _V \rho r^2 dV\ , 
U= 3 \int _V P dV \ ,
K= \int _V \rho v^2 dV \ ,
W= -\int _V \rho \mbox{\boldmath $r$} \cdot \nabla \Psi dV \ ,
$$
$$
S= -\int _S \left[P\mbox{\boldmath $r$}+ 
\mbox{\boldmath $r$}\cdot (\rho \mbox{\boldmath $v$} 
\mbox{\boldmath $v$})\right] \cdot d\mbox{\boldmath $S$} \ ,
M= {1 \over 4\pi}\int _V B^2 dV \ ,
F= \int _S \mbox{\boldmath $r$} \cdot \mbox{\boldmath $T$}_M \cdot d
\mbox{\boldmath $S$}  \ .
$$
%\end{eqnarray}
Here, $\mbox{\boldmath $T$}_M$ is the Maxwell stress-energy tensor
\begin{equation}
\mbox{\boldmath $T$}_M={1 \over 4\pi}\left(
\mbox{\boldmath $B$}\mbox{\boldmath $B$}-{1 \over 2} B^2 
\mbox{\boldmath $I$}\right) \  ,
\end{equation}
where $\mbox{\boldmath $I$}$ is the unit tensor.
The terms, $I$, $U$, $K$, $W$, $S$, $M$, and $F$ denote, respectively,  
the moment of inertia, internal thermal energy, internal kinetic
energy, gravitational energy, the sum of the thermal surface
pressure and dynamical surface pressure, internal magnetic energy,
and the magnetic surface pressure.
The internal thermal, kinetic, and magnetic energies are always
positive. The gravitational term is negative for all the 
cores in our sample, although it can in principle be positive, 
especially in crowded environments. Other terms can be either 
positive or negative.
The second term on the left hand side of equation (\ref{eq:virial})
denotes the time derivative of the flux of moment of inertia
through the boundary of the core. 
As is the standard practice (e.g., Tilley \& Pudritz 2007), we 
ignore the left hand side of equation (\ref{eq:virial}) in our 
discussion, and consider a core to be in virial equilibrium if 
the sum $U+K+W+S+M+F=0$. 

The equilibrium line is shown in Fig.~\ref{fig:virialparamnew}, where 
the sum of the surface terms ($S+F$) is plotted against 
the gravitational term ($W$), 
both normalized to the sum of the internal terms ($U+K+M$). 
The surface terms (dominated by the external kinetic term) 
are generally important; they are more so for lower mass cores 
than higher mass cores. 
The majority of the cores more massive than 1.5 times
the average core mass lie below the equilibrium line, and 
are thus bound; they are expected to collapse and form stars. 
The majority of lower mass cores lie, in contrast, above the line, 
and may disperse away, if they do not gain more mass through
accretion and/or merging with other cores, or further reduce 
their turbulence support through dissipation and/or magnetic
support through ambipolar diffusion. 

Tilley \& Pudritz (2007) found in their ideal MHD simulations 
that as the mean magnetic field becomes stronger, the surface 
terms contribute more to the virial equation.
Indeed, for their most strongly magnetized cloud, a good
fraction of cores have surface terms that are greater than 
the sum of the internal terms ($U+K+M$). This is not the 
case for our non-ideal MHD model, even though our clouds are 
more strongly magnetized than theirs. The reason is that the 
cloud dynamics is modified by ambipolar diffusion, especially 
in dense cores, where the diffusion rate is the highest. A 
related difference is that there are more bound cores in our 
models than in their strong field models, where most of the 
cores are magnetically subcritical (see also Dib et al. 2007).   
Ambipolar diffusion enables these cores to become supercritical, 
and thus more strongly bound. It is a crucial ingredient that 
cannot be ignored for core formation, even for relatively 
weakly magnetized clouds (of dimensionless mass-to-flux ratio 
$\lambda$ of a few). For such clouds, the cores tend to be more 
strongly magnetized (relative to their masses) than the cloud 
as a whole, especially in driven turbulence (Dib et al/ 2007), 
since only a fraction of the cloud mass along any given flux 
tube is compressed into a core. 

The boundedness of a core is often estimated using the virial 
parameter, the ratio of the virial mass to core mass, 
particularly in observational studies. 
In Figure \ref{fig:virialrationew}, we plot against core mass 
the virial parameter, defined as 
\begin{equation}
\alpha_{\rm vir}={5 \Delta V_{1D}^2 R_c \over GM_c},
\end{equation}
where $\Delta V_{1D}$ is the one-dimensional velocity dispersion 
including the contribution from thermal motion.
There is a clear trend for more massive cores to have smaller 
virial parameters. Indeed, nearly all cores 
more massive than 1.5 times the average 
($M_c \gtrsim 1.5 \left<M_c\right>\simeq 1.4 M_{J,s}$)
have virial parameters smaller than unity. 
It is consistent with the fact that most of these 
cores lie below the line of the virial equilibrium in 
Fig.~\ref{fig:virialparamnew},  indicating that they are bound. 
The reason for the agreement is that the surface terms are not 
as significant as the gravitational term for massive cores, and
that their magnetic support is greatly reduced by ambipolar 
diffusion (consistent with the trend for cores with smaller 
flux-to-mass ratios to have smaller virial parameters, see
Fig.~\ref{fig:virialparamnew}).
The reduction of surface terms makes the internal kinetic energy, 
which is directly related to the virial parameter, more important.
We conclude that the virial parameter can be used to gauge the 
importance of self-gravity for dense cores formed in strongly 
magnetized clouds, as long as ambipolar diffusion is treated 
properly. The virial parameter can be fitted by a power law 
$\alpha_{\rm vir} \propto M_c^{-2/3}$, in good agreement 
with the scaling found by Bertoldi \& McKee (1992) for nearby 
GMCs.

\subsection{Velocity Dispersion-Radius Relation}

The kinematics of molecular clouds and dense cores embedded in them 
are probed with molecular lines. Since the pioneering work of Larson 
(1981), many observational studies have shown that the linewidth of 
a region correlates with the size of the region. The linewidth-size 
relation is often approximated by a power law $\Delta V\propto 
R^{0.5}$. However, this relation may not hold universally, especially
on small scales. There are many cases where no clear correlation is 
found between the two (e.g., Caselli \& Myers 1995; Lada et al. 1991; 
Onishi et al. 2002).

In Fig.~\ref{fig:velocitysizenew}, we plot the 
three-dimensional
nonthermal velocity 
dispersion against radius (equivalent to the linewidth-size relation) 
for the cores in our sample. No correlation 
is apparent between the two, consistent with the observations of 
Onishi et al. (2002) for the dense cores in the Taurus molecular 
clouds. There is a large scattering in the velocity dispersion-radius 
plot, especially for cores with small masses. The reason may be that 
some cores (especially the less massive ones) are not in virial 
equilibrium and that their formation is strongly influenced by 
external flows. 

\subsection{Angular Momentum Distribution}

Angular momentum is one of the crucial parameters that determine the 
characteristics of formed stars or stellar systems. In Fig.~\ref{am1}, 
we plot the specific angular momenta of the cores against their radii.
There is no tight correlation between the specific angular momentum 
and radius, although there is a slight tendency for more massive 
cores to have somewhat larger specific angular momenta. For our 
sample cores, the specific angular momentum ranges from 0.01 to 
0.4 $c_s L_s$ (where $c_s$ is the isothermal sound speed and $L_s$ the
Jeans length of the Spitzer sheet). The distribution peaks around 
0.08 $c_s L$, corresponding 
to $0.9 \times 10^{21}$ cm$^2$ s$^{-1}$ (for the fiducial cloud 
parameters). For comparison, we 
superpose in the figure the specific angular momenta of 7 starless 
N$_2$H$^+$ cores deduced by Caselli et al. (2002). They are broadly 
consistent with our results. The average for the N$_2$H$^+$ cores 
is estimated at $\sim 1.6 \times 10^{21}$ cm$^2$ s$^{-1}$, somewhat 
larger than the peak value of our distribution. 

An oft-used quantity for characterizing the rotation of a dense core 
is the ratio of rotational energy to gravitational energy. For a 
uniform sphere of density $\rho$ rotating rigidly at an angular 
speed $\Omega$, it is given by $\beta=\Omega^2/(4\pi G\rho)$. 
Even though our cores are not uniform, and do not rotate strictly 
as a solid body, we have computed the value of $\beta$ using the 
average angular speed and mean density. The distribution of  
$\beta$ peaks around $\sim $ 0.02, in good agreement with 
the average value of 0.018 for the starless N$_2$H$^+$ cores of 
Caselli et al. (2002). It is also consistent with the well-known 
result that rotation is generally not important as far as core 
support against self-gravity is concerned (Goodman et al. 1993).  

\subsection{Flux-to-Mass Ratio}

Ambipolar diffusion enables dense cores to be less magnetized relative
to their masses. In Fig.~\ref{fig:mass2fluxnew}, we plot the magnetic 
flux-to-mass ratios of the cores against their column densities.
The ratio is normalized to the critical value $2\pi G^{1/2}$, and the 
column density to that of the initially uniform state, 
$4\pi\rho_0 L_s$. 
The magnetic flux-to-mass ratio is 
approximated as $\Gamma_c\equiv \pi R_{c}^2 \left< B\right>/M_c$, where
$\left<B\right>$ is the mean magnetic field strength inside a core
(e.g., Dib et al. 2007) and the mean core radius $R_c$ is 
defined as the radius of a sphere enclosing the same volume as 
the core. The column density is calculated from the core mass 
divided by the area of a circle whose radius is equal to $R_{c}$. 
There is a clear trend for cores with higher column densities to have 
lower flux-to-mass ratios. Indeed, all cores with masses greater 
than $1.5\left<M_c\right>$ are magnetically supercritical, with 
flux-to-mass ratios as low as $\sim 0.35$, indicating 
that they have all experienced significant ambipolar diffusion. 
The weakening of magnetic support may have enabled these cores to 
condense to a high column density in the first place. 

\subsection{Core Mass Spectrum}

In Figure \ref{fig:coremf} we plot the mass spectrum for our sample 
of dense cores. The core mass is normalized to the Jeans mass of 
the Spitzer sheet, $M_{J,s}$. 
The lowest mass at $\sim 0.1~M_{J,s}$ corresponds to the minimum 
core mass determined from the criteria for core identification. There 
is a prominent break around $1~M_{J,s}$ in the mass spectrum. Above 
the break, the spectrum can be fitted roughly by a power law $dN/dM 
\propto M^{-2.5}$, not far from the Salpeter stellar initial mass 
function. We note that cores above the break contain more than 
$\sim 100$ cells; their structures should be reasonably well resolved. 
The core spectrum flattens below the break, which is also
broadly consistent with observations. 
Similar core spectra are also 
obtained in the turbulent fragmentation model of core formation that
involves much weaker magnetic field 
and much stronger turbulence
(e.g., Padoan \& Nordlund 2002).
The characteristic mass below which the core mass spectrum flattens 
apparently coincides with the Jeans mass of the condensed sheet, 
which may result from magnetically regulated gravitational fragmentation.
Caution must be exercised in trying to differentiate various 
scenarios of core formation, such as the magnetically regulated 
gravitational fragmentation and the turbulent fragmentation, based 
on core mass spectrum and, perhaps by extension, stellar mass 
spectrum. The stellar mass spectrum for the 52 stars formed in the 
standard model and its close variant are plotted in Fig.~\ref{fig:coremf}
for comparison. The stellar mass spectrum also resembles the 
Salpeter IMF. 

\section{Discussion}
\label{discussion}

\subsection{Magnetically-Regulated Star Formation in Quiescent 
Condensations} 
\label{scenario}

We have simulated star formation in relatively diffuse, mildly 
magnetically subcritical clouds that are strongly turbulent 
initially. The decay of initial turbulence leads to gravitational 
condensation of the diffuse material along field lines into 
a highly fragmented sheet (see Fig.~\ref{3D}), which produces 
stars in small groups. The bulk of the condensed material 
remains magnetically supported in the cross-field direction, 
however, with typical motions that are moderately supersonic 
and approximately magnetosonic. Only a small fraction of the 
condensed material collapses gravitationally to form stars. 
The calculations lead us to a picture of slow, inefficient, 
star formation in a relatively quiescent, large-scale 
condensation (which can be a filament if the cloud is flattened 
in a direction perpendicular to the field lines to begin with) 
that is embedded in a much more turbulent halo of diffuse gas. 

The above picture is an unavoidable consequence of the anisotropy 
intrinsic to the large-scale magnetic support. Before the self-gravity 
becomes strong enough to overwhelm the magnetic support in the 
cross-field direction (to initiate widespread star formation), it  
should be able to condense material along the field lines (where 
there is no magnetic resistance) first. The condensed material 
is expected to be relatively quiescent. Its field-aligned velocity 
component is damped during the condensation process, whereas its 
cross-field motions are constrained by the field lines. The 
lateral magnetic support ensures that the bulk of the condensed 
material is relatively long-lived, allowing more time for the 
turbulence to dissipate. The dissipation is hastened by the high 
density in the condensation, which makes it easier for shock 
formation, because of low Alfv\'en speed. Indeed, when the condensed 
layer accumulates enough material for its mass-to-flux 
ratio to approach the critical value (a condition for widespread 
star formation), its Alfv\'en speed drops automatically to a value 
comparable to the sound speed (NL05, Kudoh et al. 2007). Rapid 
dissipation of super-magnetosonic motions ensures that the bulk of 
the condensed material moves at no more than moderately supersonic 
speeds, as we find here and previously in 2D calculations (NL05). 
It is in this relatively quiescent environment that most star 
formation takes place.

The quiescent, star-forming, condensed material is surrounded by 
a highly turbulent, fluffy halo. The stratification in turbulent 
speed is a consequence of the stratification in density (Kudoh \& 
Basu 2003): the amplitude of a wave increases as it propagates 
from high density to low density, and faster motions can survive 
for longer in the lower density region because of higher Alfv\'en 
speed. We believe that this is a generic result independent of 
the source of turbulence. In our particular simulations, the 
saturated, highly supersonic turbulence in the halo at late times 
is driven mainly by protostellar outflows internally. It is 
conceivable, perhaps even likely, that most of the turbulence 
in the diffuse regions of star forming clouds comes from energy 
cascade from larger-scale molecular gas (perhaps even HI envelope). 
Depending on the details of energy injection, the external driving 
of turbulence can slow down the 
gravitational settling along field lines and may even suppress 
it altogether. Even when the large-scale gravitational settling 
is suppressed, star formation can still occur, perhaps in a less 
coherent fashion: stars can form in localized pockets of a 
magnetically subcritical cloud directly from strong compression 
through shock-enhanced ambipolar diffusion (NL05), especially
when the shock is highly super-Alfv\'enic (Kudoh \& Basu 2008). 
Vigorous driving may favor a mode of turbulence-accelerated star 
formation at isolated locations, whereas decay of turbulence may 
allow the gravity to pull the diffuse gas into large-scale 
condensations, in which stars form in a more coherent fashion. 
The first mode is now under investigation, and the results will 
be reported elsewhere. The second corresponds to the picture 
outlined above.  

The dichotomy is not unique to star formation in strongly magnetized 
clouds. It is also present when the magnetic field is weak or 
non-existent: whereas stars form rapidly in a clustered mode in a 
decaying non-magnetic turbulence, they are produced more slowly, in 
a more dispersed fashion, if the turbulence is driven constantly, 
especially on small scales (e.g., Mac Low \& Klessen 2004). Different 
modes of star formation may operate in different environments, and 
there is a urgent need for sorting out which mode dominates in which 
environment. In the next subsection, we will concentrate on the Taurus 
molecular clouds, the prototype of distributed low-mass star forming 
regions (see Kenyon et al. 2008 for a recent review), and the Pipe 
nebula, which appears to be a younger version of the Taurus complex 
(Lada et al. 2008). 

\subsection{Connection to Observations} 

\subsubsection{The Taurus Molecular Cloud Complex}
\label{taurus}

The Taurus molecular cloud complex may represent the best case for 
the picture of magnetically regulated star formation investigated 
in this paper. One line of evidence supporting this assertion 
comes from the studies of young stars by Palla \& Stahler (2002). 
They inferred, using pre-main sequence tracks, that the star formation 
in the complex has two phases. It began at least 10 Myrs ago, at 
a relatively low level and in widely dispersed locations (Phase 
I hereafter). In the last three million years or so, stars are 
produced at a much higher rate (Phase II). The majority of the 
stars produced in Phase II are confined within the dense, 
striated, C$^{18}$O filaments (Onishi et al. 2002). Palla \& 
Stahler interpret these results to mean that the Taurus clouds 
evolved from an initial, relatively diffuse, state 
quasi-statically. They envision that, during the early epoch 
of quasi-static contraction, few places have contracted to 
the point of forming dense cores and protostars, and hence only 
scattered star formation activity is seen at isolated locations 
(Phase I). 
Eventually, further contraction of the cloud leads to the formation 
of nearly contiguous, dense filaments where multiple cores can 
form and collapse simultaneously, leading to elevated star formation 
(Phase II). 
% The authors did not speculate on the physical mechanisms 
% that may be responsible for this behavior. 
In view of the discussion 
in the last subsection, it is natural to identify Phase I with the 
shock-accelerated mode of star formation in a magnetically 
subcritical or nearly critical cloud (or sub-region), when the 
turbulence is still 
strong enough to prevent large-scale gravitational condensation 
along the field lines. In this picture, Phase II started when the
turbulence had decayed to a low enough level that the gravity was  
able to collect enough mass along the field lines into condensed 
structures for a faster, more organized, mode of star formation 
to take over. 

A salient feature of our picture is that the rate of star formation 
in Phase II should remain well below the limiting free-fall rate 
of the condensed material. There is strong evidence that the 
conversion of dense material in the Taurus clouds into stars is 
indeed slow. The dense, filamentary material is well traced by 
C$^{18}$O, which Onishi et al. (1996) has studied in great detail. 
These authors estimated a mass and average density for the C$^{18}$O 
gas of $M_{{\rm C}^{18}{\rm O}}\sim 2900~M_\odot$ and $n_{H_2}\sim
4000$~cm$^{-3}$ (comparable to that in the condensed sheet in 
our simulations). The latter corresponds a free fall time 
$t_{{\rm C}^{18}{\rm O},ff}=4.86 \times 10^5$~yrs. The maximum, free-fall, 
rate of turning C$^{18}$O material into stars is therefore 
${\dot M}_{*,{\rm C}^{18}{\rm O},ff}\sim 6\times
10^{-3}~M_\odot$~yr$^{-1}$. This 
rate is to be compared with the actual rate of star formation in 
the complex, ${\dot M}_*\sim 5\times 10^{-5}~M_\odot$yr$^{-1}$, 
estimated by Goldsmith et al. (2008) based on an assumed average 
mass of $0.6~M_\odot$ for the 230 stars observed in their
mapped region (Kenyon et al. 2008) and a duration of 3~Myrs for 
Phase II of star formation where most stars are produced (Palla 
\& Stahler 2002). Therefore, the rate of converting the even 
dense (relative to the diffuse $^{12}$CO gas) C$^{18}$O material 
into stars is extremely slow, amounting to $\sim 1\%$ per local 
free fall time. This rate is comparable to those obtained in our 
simulations. 
%
% comparable to Krumholz's estimates for denser gas!
% 

One may argue that the slow rate of star formation in C$^{18}$O
gas is due to turbulence support. However, the C$^{18}$O gas 
in the Taurus clouds is observed to be unusually quiescent. 
Onishi et al. (1996) estimated a mean equivalent line-width 
(defined as the integrated intensity divided by the peak 
temperature of the line) of 0.65~km~s$^{-1}$, corresponding 
to a 3D velocity dispersion of 0.45~km~s$^{-1}$ for a Gaussian
profile. The velocity dispersion is $\sim 2.4$ times the 
sound speed for a gas temperature of 10~K, indicating that 
the moderately dense gas traced by C$^{18}$O is not 
highly supersonic in general. Indeed, the velocity 
dispersion is quite comparable to that of the condensed
sheet material in our standard model at similar densities
(see Fig.~\ref{rms_vel}). Nearly 1/3 ($937~M_\odot$ out 
of $2900~M_\odot$) of the C$^{18}$O gas is classified 
into 40 cores, which have a mean line-width (FWHM) of 
0.49~km~s$^{-1}$, comparable to that estimated above 
for the C$^{18}$O gas as a whole. Onishi et al. concluded 
that most of the cores are roughly gravitationally bound, 
based on their masses, sizes and line-widths. These authors 
argued that the general alignment of the minor axes of the 
cores with the overall field direction indicates that the 
magnetic field plays a role in their formation. We agree 
with this assessment. 
The relatively low velocity dispersion of the C$^{18}$O gas 
and the morphology of dense cores provide indirect support 
for the picture of magnetically regulated star formation. 

A more direct test of the picture would come from a 
determination of the mass-to-flux ratio. 
%
% what is predicted? can moderately supercritical clouds 
% work? 
% 
For a given field strength $B$, the critical column density 
along the field line is
\begin{equation}
\Sigma_{\vert\vert,c}=58.7~B_{20}~M_\odot~{\rm pc}^{-2}, 
\label{critical}
\end{equation}
corresponding to an $A_V \approx 2.5~B_{20}$, where $B_{20}$ 
is the field strength in units of 20~$\mu$G. 
Goldsmith et al. (2008) estimated a total mass of 
$2.4\times 10^4 M_\odot$ in their 
mapped area of $28$~pc$\times 21$~pc, yielding an average 
column density of 40.8~M$_\odot$~pc$^{-2}$ along the line 
of sight. Only $\sim 2/3$ of the surveyed area have detectable 
$^{12}$CO emission in individual spectra, however (their 
Masks 1 and 2). For these regions, the total mass and area 
are $1.95\times 10^4 M_\odot$ and $\sim 392$~pc$^2$, yielding 
a somewhat higher average column density of 49.7~M$_\odot$~pc$^{-2}$ 
along the line of sight. If the column density along the field 
lines is comparable to this value, the corresponding critical 
field strength would be $16.9~\mu$G, according to 
equation~(\ref{critical}). About half of the mass (9807~M$_\odot$) 
resides in eight ``high-density'' regions (such as L1495 and 
Heiles Cloud 2) that together occupy $\sim 31\%$ of the area, 
with an average line-of-sight column density of 
79.5~M$_\odot$~pc$^{-2}$. This value would correspond to 
a critical field strength of $27.1~\mu$G if the magnetic field 
is exactly along the line of sight. There is, however, evidence 
for a large-scale magnetic field in the plane of sky in the 
Taurus cloud complex from the polarization of background 
star light. Indeed, the velocity anisotropy measured by Heyer 
et al. (2008) in a relatively diffuse sub-region points to 
a magnetic field that is mainly in the plane of sky. If this 
magnetic orientation is correct, and if matter condenses along the 
field lines into flattened structures, then projection effects 
may be considerable. Applying a moderate correction factor 
of $\sqrt{2}$ (corresponding to a viewing angle of $45^\circ$) 
would bring the average field-aligned column density of the 
``high-density'' regions down to 56.2~M$_\odot$~pc$^{-2}$, with 
a corresponding  
critical field strength of $19.2~\mu$~G. It is therefore likely 
that a magnetic field of order 20~$\mu$G is strong enough to 
regulate star formation in the Taurus cloud complex. 

The magnetic field strength has been measured for three dense 
cores in Taurus through OH Zeeman measurements using Arecibo 
telescope (Troland \& Crutcher 2008). They are B217-2, TMC-1 
and L1544, with a line-of-sight field strength of $B_{\rm los} 
\approx 13.5$, $9.1$, and $10.8$~$\mu$G respectively. Although 
the true orientation of the magnetic field is unknown, a 
statistically most probable correction factor of 2 would bring 
the total field strength to $\sim 18.2 - 27.0~\mu$G. Given that 
OH samples moderately dense gas approximately as C$^{18}$O 
does, and that the field strength changes relatively little at 
low to moderate densities (see Fig.~\ref{ave_B}), it is plausible 
such a field permeates most of the space in the complex. A 
worry is that such a strong field is not detected in several 
other cores in the Taurus in the same Arecibo survey. Whether
these non-detections can be reconciled with our picture of
magnetically regulated star formation through, e.g., unfavorable 
field orientation, remains to be seen. 

Indirect evidence for a relatively strong magnetic field 
comes from the observation that the magnetic field as
traced by the polarization vectors of the background star
light is well ordered over many degrees in the sky, 
indicating that the magnetic energy is larger than the 
turbulent energy.
If we use conservative estimates for H$_2$ number density 
of $10^2$~cm$^{-3}$ (see Table 2 of Goldsmith et al. 2008), 
and turbulent speed of 2~km~s$^{-1}$,
we can put a lower limit to the field strength at $\sim 
15~\mu$G. The limit agrees with the value of $\sim 14~\mu$G 
estimated by Heyer et al. (2008) based on comparing the 
observed velocity anisotropy in a striated, relatively 
diffuse region with MHD turbulence simulations; they point 
out that 
the value is to be increased if the magnetic field has 
a substantial component along the line of sight. A
strong line-of-sight magnetic field of $\sim 20~\mu$G 
or more is deduced by Wolleben \& Reich (2004) at the 
boundaries of molecular clouds a few degrees away from 
the sub-region studied by Heyer et al. (2008), from 
modeling the enhanced rotation measures inferred from 
polarization observations at 21 and 18~cm. Subsequent 
HI Zeeman observations did not confirm the deduced 
field strength, however (C. Heiles, priv. comm.). 

Although none of the pieces of evidence discussed above is 
conclusive individually, taken together, they make a 
reasonably strong case for a global magnetic field of order 
$\sim 20\mu$G in strength. The existence of such a field 
may not be too surprising from an evolutionary point of view. 
It is now established that the cold neutral medium (CNM) of 
HI gas (likely precursor of Taurus-like molecular clouds)   
is fairly strongly magnetized in general, with a median
field strength of $\sim 6~\mu$G (Heiles \& Troland 2005). 
For denser HI clouds, it is not difficult to imagine a 
somewhat higher field strength. 
A case in point is the nearby Riegel-Crutcher HI cloud 
that we mentioned in the introduction. It was mapped 
recently by McClure-Griffiths et al. (2006) using 21~cm 
absorption against strong continuum emission towards the 
Galactic center. They found long strands of cold neutral
hydrogen that are remarkably similar to the CO striations 
in Taurus studied by Heyer et al. (2008): both are aligned 
with the local magnetic field. For the magnetic energy to
dominate the turbulent energy inside the filaments, the
field strength must be of order $30~\mu$G or larger. This
value is consistent with $B_{los}\sim 18~\mu$G obtained 
by Kazes \& Crutcher (1986) through Zeeman measurements 
at a nearby location, if a correction of $\sim 2$ is
applied for projection effects. The physical conditions 
inside the HI filaments of the Riegel-Crutcher cloud are 
not dissimilar to those of diffuse CO clouds: number 
density $n_H\sim 460$~cm$^{-3}$, 
velocity dispersion $\sigma \sim 3.5$~km~s$^{-1}$, and 
temperature $T\sim 40$~K. The main difference is that the 
column density is $\sim 10^{20}$~cm$^{-2}$, which is not 
sufficient for shielding the CO from photodissociation 
(van Dishoeck \& Black 1988). 

The above discussion leads us to the following scenario for 
cloud evolution and star formation in the Taurus molecular
clouds: we envision the cloud complex to have evolved from a 
turbulent, strongly magnetized, non-self-gravitating HI cloud 
similar to the Riegel-Crutcher cloud. Once enough material 
has been accumulated, perhaps by converging flows 
(Ballesteros-Paredes et al. 1999) or some other means
(such as differential turbulence dissipation, which may 
lead to a condensation through ``cooling flow'' along the 
field lines, e.g., Myers \& Lazarian 1998), for CO 
shielding, it becomes visible as a molecular cloud. 
Depending on the degree of magnetization, level of 
inhomogeneity, and strength of turbulence, star formation 
can start quickly in some pockets (perhaps even during the 
process of molecular cloud formation), despite of a strong 
global magnetic field, either because they happen to be 
less magnetized to begin with or because their magnetic 
fluxes are force-reduced by shock-enhanced ambipolar 
diffusion. This corresponds to the low-level, Phase I of 
star formation deduced by Palla \& Stahler (2002). Once 
enough material has settled along field lines for the  
mass-to-flux ratio of the condensed material to approach 
the critical value, the rate of star formation increases 
drastically. We identify this phase of more organized star 
formation in condensed structures as the more active, 
Phase II of star formation deduced by Palla \& Stahler 
(2002). Even during this relatively active phase, the star 
formation rate is still well below the free-fall rate of 
the condensed gas, because of magnetic regulation. 
%
% 330 solar masses in H13CO+, SFE rate 10 times slower than free 
% fall rate?? 
%

The above picture can be contrasted with that of Ballesteros-Paredes 
et al. (1999) and Hartmann et al. (2001). These authors advocated 
a picture of rapid cloud formation and star formation for the
Taurus-Auriga complex, motivated mainly by a lack of ``post-T Tauri 
stars'' older than $\sim 5$~Myrs in the region; indeed, the 
majority of stars are younger than $\sim 3$~Myrs, as mentioned 
earlier. They argue that it is difficult to understand how star 
formation can occur simultaneously over a spatial extent of $\sim 
20$~pc (or more) over a timescale as short as a few million years, 
given that the observed turbulent speed is $\sim 2$~km~s$^{-1}$, 
corresponding to a crossing time of $\sim 10$~Myrs (or more), 
unless the star formation is coordinated by some agent other than 
the currently observed turbulence. They suggest that this agent is 
external: a large-scale converging flow with speeds of 5-10~km~s$^{-1}$.  
In our picture, it is the global gravity and a strong, large-scale, 
internal magnetic field that coordinate the star formation: the 
strong field suppresses vigorous star formation until enough 
material has collected along field lines into condensed structures 
for the gravity to start overwhelm magnetic support in the 
cross-field directions. The degree to which the star formation is 
actually synchronized will, of course, depend on how the mass is 
initially distributed along the field lines and rate of turbulence 
dissipation and replenishment, neither of which is, unfortunately, 
well constrained at the present. Indeed, a hybrid scenario may be
possible, with the mass accumulation along the field lines sped 
up by a large-scale converging flow. Nevertheless, we believe 
that, at least in the particular case of the Taurus molecular cloud 
complex, for the large amount of dense, relatively quiescent, 
C$^{18}$O gas to form stars at $\sim 1\%$ of the free-fall rate, 
magnetic regulation is needed. A similar case can be made for the 
Pipe Nebula.   

\subsubsection{Pipe Nebula}
\label{pipe}

Although less well studied than the Taurus cloud complex, the Pipe 
nebula is an important testing ground for theories of low-mass 
star formation in relatively diffuse dark clouds. It has $\sim 
10^4~M_\odot$ of molecular material as traced by $^{12}$CO (Onishi 
et al. 1999), not much different from the Taurus complex. The higher 
density gas traced by $^{13}$CO and C$^{18}$O is concentrated in 
a long filament that spans more than 10~pc, again similar to the
Taurus case. The main difference is that there is no evidence
for star formation activity except in the most massive core, B59. 
The lower rate of star formation indicates that the Pipe nebula is 
at an earlier stage of evolution compared to the Taurus complex, as 
pointed out by Muench et al. (2007). The relative inactivity is 
not due to a lack of dense, quiescent material that is a 
pre-requisite of (low-mass) star formation. J. Alves et al. (2007) 
identified 159 dense cores from extinction maps, many of which are 
formally gravitationally unstable, with masses above the critical 
Bonner-Ebert mass (Lada et al. 2008). In the absence of additional 
support, such cores should collapse and form stars on the free-fall 
time scale, which is $t_{ff}=3.7\times 10^5$~yrs for the inferred 
median core density $n(H_2)=7\times 10^3$~cm$^{-3}$. This time 
scale is much shorter than the turbulence crossing time $t_c=L/v 
\sim 5\times 10^6$~yrs, over a region of $L \sim 15$~pc, and for 
a typical turbulent speed of $v\sim 3$~km~s$^{-1}$. The mismatch 
between the turbulent crossing time and star formation time is 
therefore much more severe than that noted above for the Taurus 
region. 

There are two possible resolutions to the above conundrum. If 
the dense cores (at least those massive enough to be formally 
gravitationally unstable) are indeed short-lived objects, with 
a lifetime of order the average free-fall time ($\sim 3\times 
10^5$~yrs), their formation must be rapid and well coordinated. 
It is tempting to 
attribute the coordination to a large-scale converging flow, as 
Ballesteros-Paredes et al. (1999) and Hartmann et al. (2001) 
did for the Taurus case. However, to cover a distance of $15$~pc 
(the size of the cloud, which is comparable to the length of 
the chain of dense cores) in $\sim 3\times 10^5$ years, the 
flow needs to converge at a speed of order $\sim 50$~km~$s^{-1}$, 
which is uncomfortably high. A more likely alternative is that 
the cores live much longer 
than their free-fall times would indicate. While the longevity 
is not necessarily a problem for those non-self-gravitating, 
lower mass cores that can be confined by external pressure, 
it is problematic for the more massive cores that should have 
collapsed on the free-fall time scale, unless they are supported 
by some (so-far invisible) agent in addition to the observed
thermal and turbulent pressures. The most likely candidate is
a magnetic field. There is some evidence for the existence of 
an ordered magnetic field that is more or less perpendicular 
to the long filament that contains the dense cores from stellar 
polarization observations (F. Alves \& Franco 2007; F. Alves et
al., in preparation), although much more work is needed, particularly 
for the diffuse region outside the filament, to firm up or reject 
this possibility. In any case, it is plausible that 
the Pipe nebula represents an earlier stage of the magnetically 
regulated cloud evolution and star formation than the Taurus 
clouds: material has started to collect along the field lines 
into dense, relatively quiescent, structures, but their 
mass-to-flux ratios are yet to reach the critical value, 
except perhaps in the most massive core, B59. In this case, 
the ages of the cores are set by the rate of mass condensation 
along the field lines, which may depend on the ill-understood 
process of turbulence replenishment in the diffuse gas, and 
may be accelerated by large-scale converging flows (Ballesteros 
et al. 1999; Hartmann et al. 2001). 
%
% Southern coalsack??? earlier phase???   
%

\subsection{Modes of Star Formation in Strongly Magnetized Clouds}
\label{mode}

Strong magnetic fields tend to resist cloud condensation and 
star formation. This resistance can be overcome by either 
turbulence or gravity, leading to two distinct modes of star 
formation. Which mode dominates depends on the relative
importance of gravity and turbulence. When the turbulence is 
strong enough to prevent gravitational settling along field 
lines, star formation is still possible. Turbulence can 
overcome magnetic resistance in localized regions that are 
compressed by converging flows to high densities; such regions 
are expected to have large magnetic field gradients and low 
fractional ionization, both of which enhance the rate of 
ambipolar diffusion. The 
turbulence-accelerated ambipolar diffusion enables dense
pockets to become self-gravitating 
%\footnote{Dib et al. (2007) 
%found that the majority of the dense cores formed in a rather 
%weakly magnetized cloud of dimensionless mass-to-flux ratio 
%of $\lambda=2.8$ with driven turbulence are nearly magnetically 
%critical or even subcritical. Such cores are expected to be
%strongly affected by enhanced ambipolar diffusion as well.} 
and collapse into stars (NL05, Kudoh \& Basu 2008), even when 
the background cloud 
remains well supported. In this mode, the crucial step of 
star formation---core formation---is driven by turbulent 
converging flows but regulated by magnetic fields, with 
self-gravity playing a secondary role. After a bound core 
has formed, its further evolution is driven primarily by 
the self-gravity, but remains regulated by the magnetic 
field. 

The second mode corresponds to a turbulence that is not strong 
enough to prevent gravitational settling, either because it 
is weak to begin with or because it has decayed without being 
adequately replenished. Condensation would occur along field 
lines, which would lead to further increase in self-gravity 
and more rapid dissipation of turbulence. In the condensed 
structure, turbulence becomes less important dynamically, and 
star formation is driven mainly by gravity in the cross-field 
direction. The gravitational contraction is resisted primarily 
by magnetic fields, although protostellar outflows play an 
important role in dispersing away the dense gas that remains 
after the formation of individual stars; the dispersal slows 
down the global star formation. In this mode, the gravity and 
magnetic fields are of primary importance, with turbulence 
playing a secondary role. 

The above two modes of star formation may be interconnected: it 
is natural to expect the second to follow the first, as a result 
of turbulence dissipation. The transition from one to the 
other should occur when the mass-to-flux ratios in the condensed 
structures approach the critical value. It may be possible for 
the mass-to-flux ratio to stay near the critical value due to 
outflow feedback: if too much mass is added to the structure 
so that its mass-to-flux ratio increases well above the critical 
value, it would form stars at a higher rate, producing more 
outflows that stop further mass accumulation, similar to the 
outflow-regulated scenario of cluster formation envisioned 
in Matzner \& McKee (2000). They have shown that regions of 
outflow-regulated star formation tend to be overstable, 
oscillating with increasingly large amplitudes. Indeed, the 
outflows may unbind the condensed structure completely, 
terminating its star formation quickly. An abrupt termination 
is implied by the dearth of the well-observed star forming 
regions in the declining phase of star formation (Palla \& 
Stahler 2000). 

If, on the other hand, outflows fail to retard mass accumulation 
along field lines, the star formation may change over to 
yet another, more active, mode --- cluster formation. The 
transition is expected to take place when the mass-to-flux ratios 
of the condensations become substantially greater than the critical 
value (say by a factor of two), making global (not just local) 
contraction perpendicular to the field lines possible. The
higher rate of star formation should lead to a stronger 
outflow feedback, which could increase the turbulence 
to a level that balances the global collapse (e.g., Nakamura 
\& Li 2007; Matzner 2007). In this mode, protostellar 
outflow-driven turbulence becomes the primary agent that 
resists the global gravitational collapse, with the magnetic 
field playing a secondary role. 

In summary, we envision three, perhaps sequential, modes of star 
formation in strongly magnetized clouds, dominated (a) first by 
turbulence and magnetic fields, (b) then by magnetic fields and 
gravity, and (c) finally by gravity and turbulence (again). In 
this sequence, stars form first directly out of turbulent diffuse 
material in a dispersed fashion, then more coherently in 
relatively quiescent, condensed structures, and finally as 
clusters in dense turbulent clumps. The strong turbulence 
is primordial initially. It is transformed into a protostellar 
turbulence by outflows in regions of active cluster formation.
%
% B important in quiescent regions, emphasized by Cox; true for SF as well. 
%
% (b) is unique to strong field case

\section{Summary and Conclusions}
\label{conclude}

\begin{figure}
\plotone{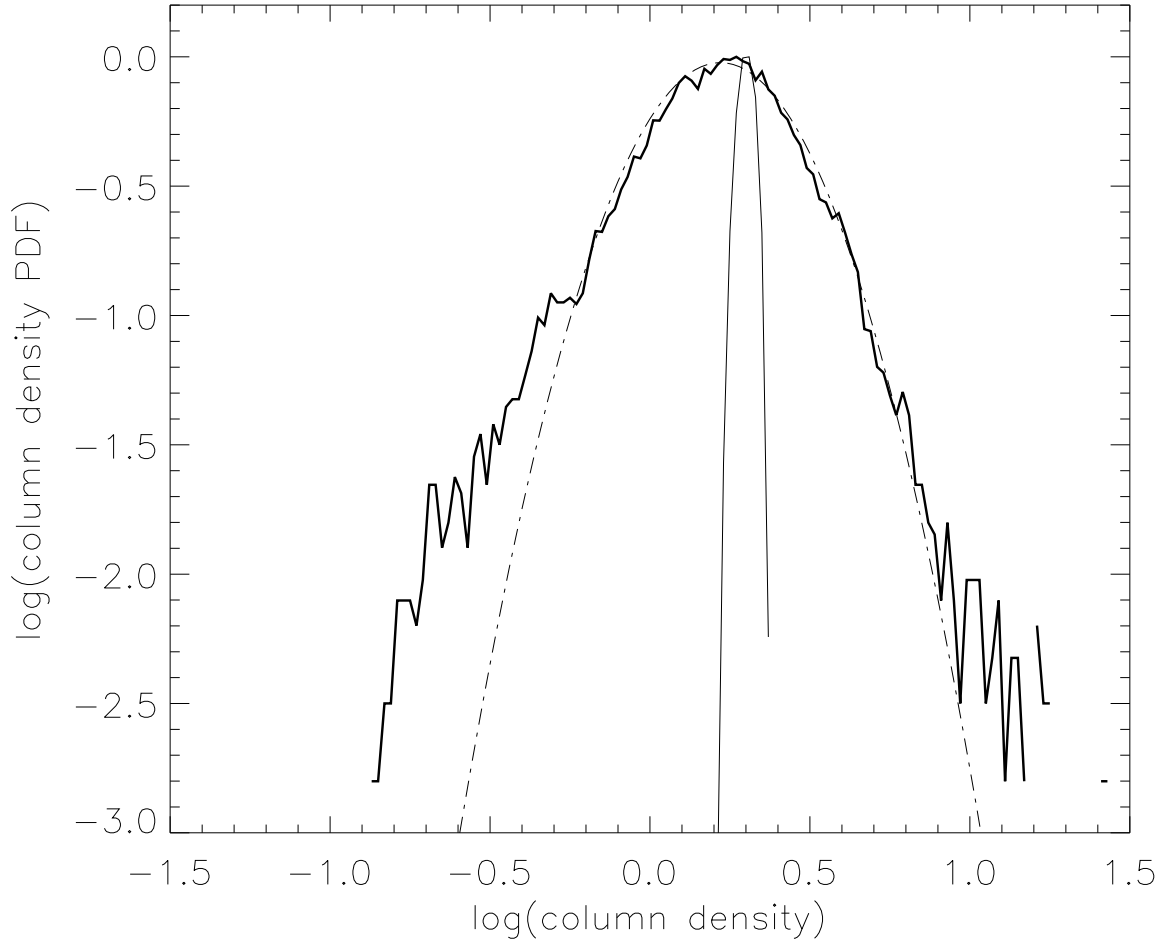}
\caption{The PDF of the face-on column density of the standard model 
shown in Fig.~\ref{colden_critical} ({\it thick solid line}), with a 
fitted lognormal distribution ({\it dashed-dotted line}) superposed.
The distribution is normalized so that the peak value is close to 
unity, and the column density is in units of $\rho_0 L_J$. Also
plotted for comparison is the column density PDF for the ideal MHD
counterpart of the standard model ({\it thin solid line}), which is 
much narrower. The stark contrast between the two PDFs highlights the 
key role of ambipolar diffusion in fragmenting strongly magnetized
clouds.}  
\label{colden_PDF}
\end{figure}

\begin{figure}
%\epsscale{0.6}
\plotone{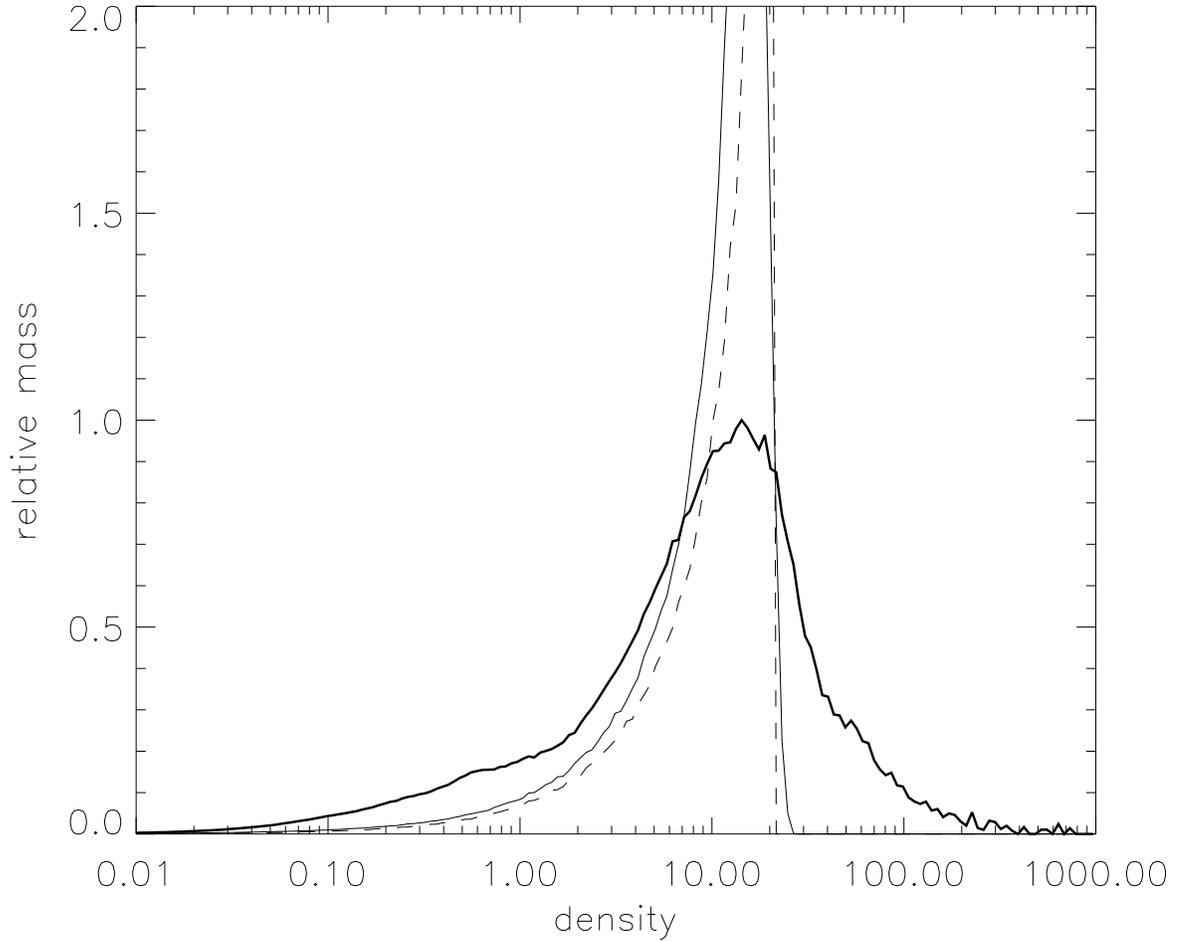}
\caption{Distribution of mass as a function of density (in units 
of initial density $\rho_0$, {\it thick solid line}), along with 
that for the ideal MHD case ({\it thin solid line}) and 
for a Spitzer sheet containing the same amount of mass ({\it 
dashed line}). The mass distribution is much broader in the 
presence of ambipolar diffusion, because of (magnetically-diluted) 
gravitational fragmentation.}  
\label{lognorm_density}
\end{figure}

\begin{figure}
%\plotone{rms_vel.eps}
%\epsscale{0.6}
\plotone{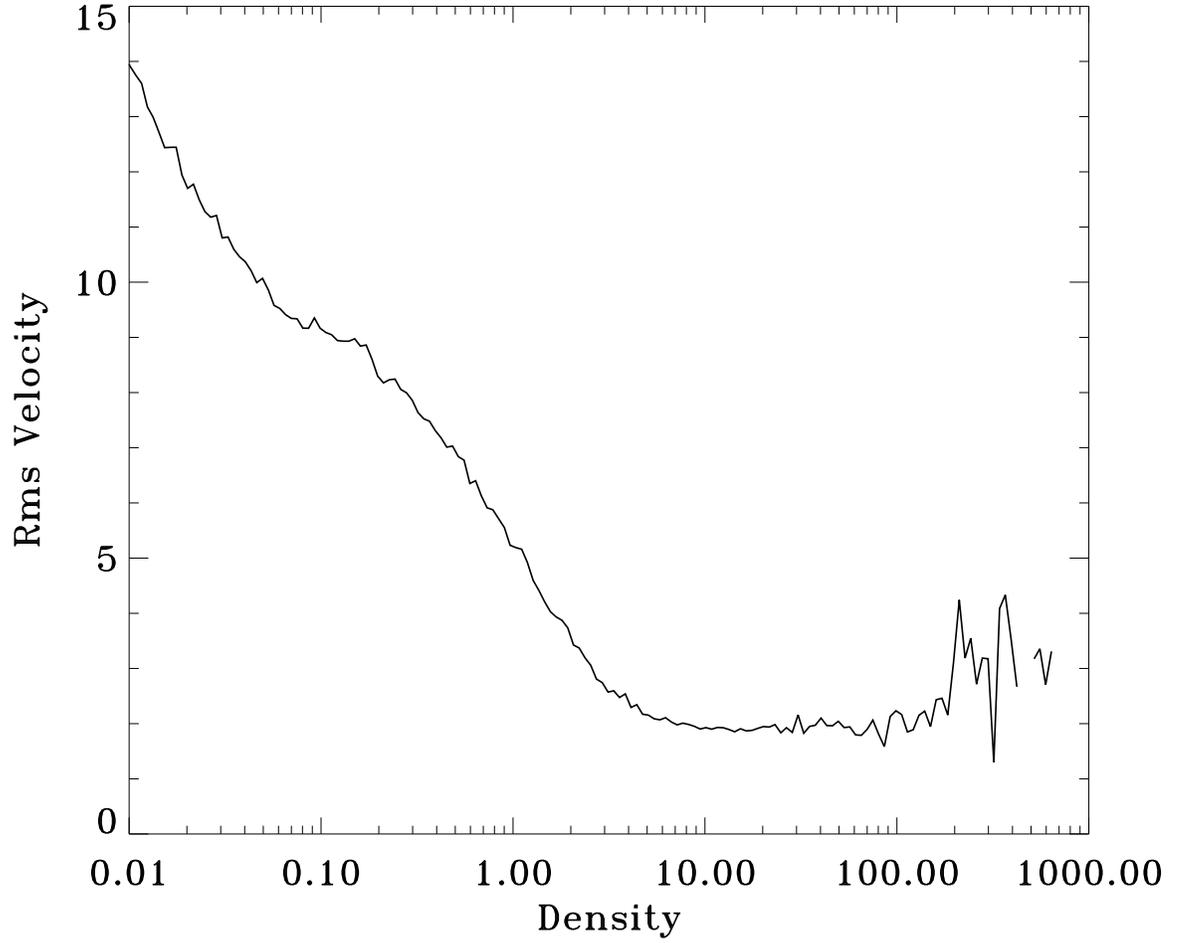}
\caption{Distribution of mass-weighted rms velocity (in units of 
sound speed $c_s$) as a function of density (in units of $\rho_0$),
showing that the diffuse halo is much more turbulent than the 
condensed sheet.}  
\label{rms_vel}
\end{figure}

\begin{figure}
%\epsscale{0.6}
\plotone{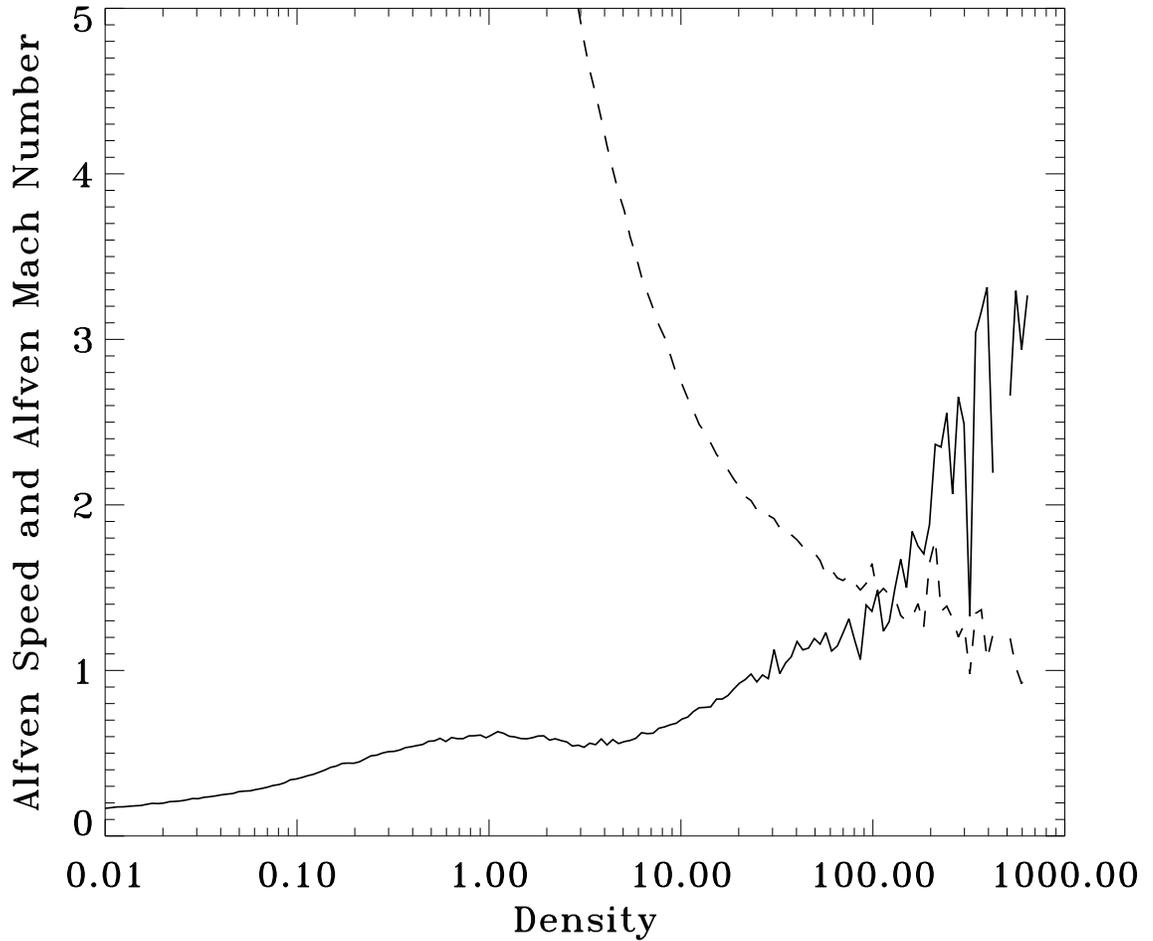}
\caption{Distribution of the average Alfv\'en Mach number $M_A$ ({\it
    solid line}) and Alfv\'en speed (in units of sound speed, {\it 
dashed line}) as a function of density (in units of $\rho_0$), showing
that the motions of the bulk condensed sheet materials are roughly
Alfv\'enic whereas those of the halo materials are sub-Alfv\'enic.} 
\label{ave_MA}
\end{figure}

\begin{figure}
%\epsscale{0.6}
\plotone{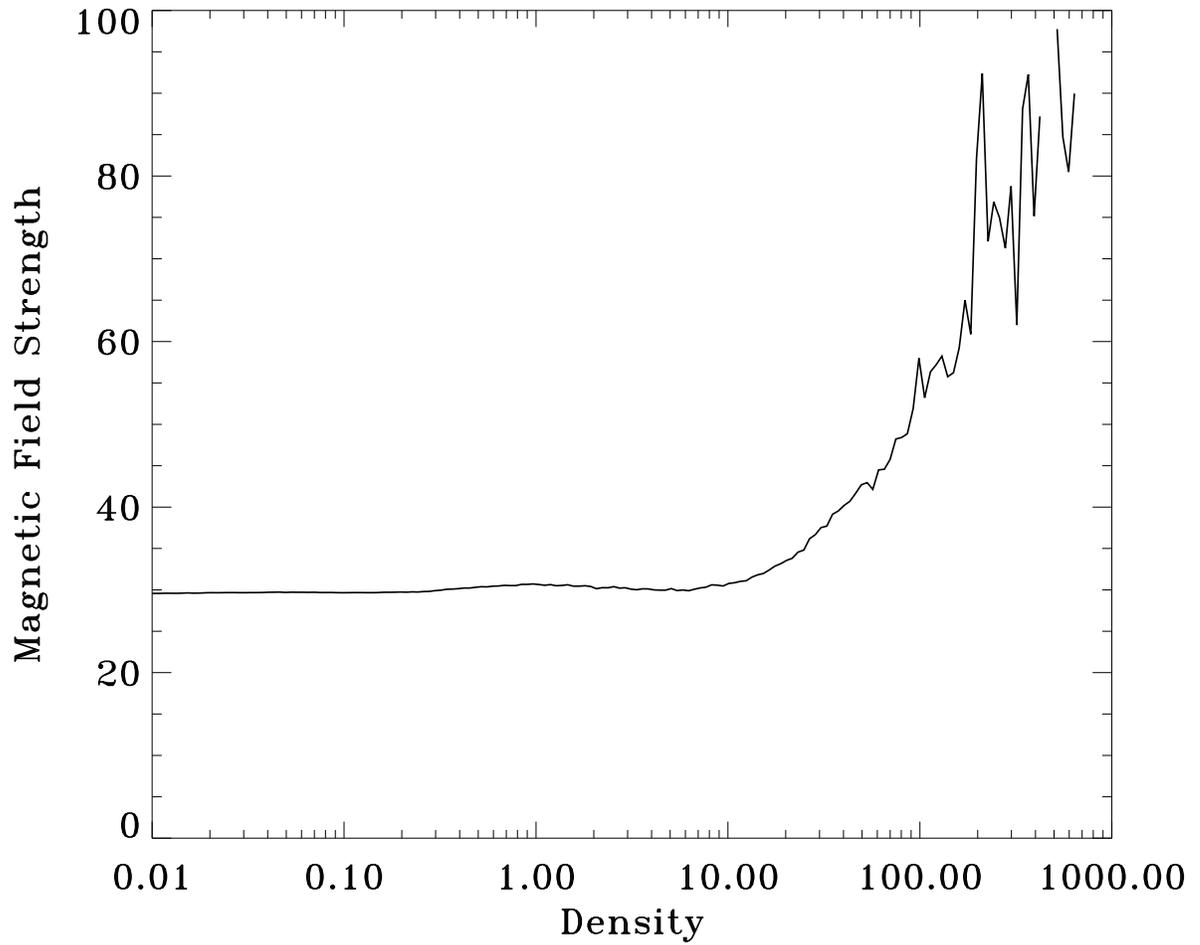}
\caption{Distribution of average magnetic field strength (in units 
of the initial $B_0$, see eq.~[\ref{fieldstrength}]) as a function 
of density (in units of $\rho_0$).} 
\label{ave_B}
\end{figure}

\begin{figure}
%\epsscale{1.0}
\plotone{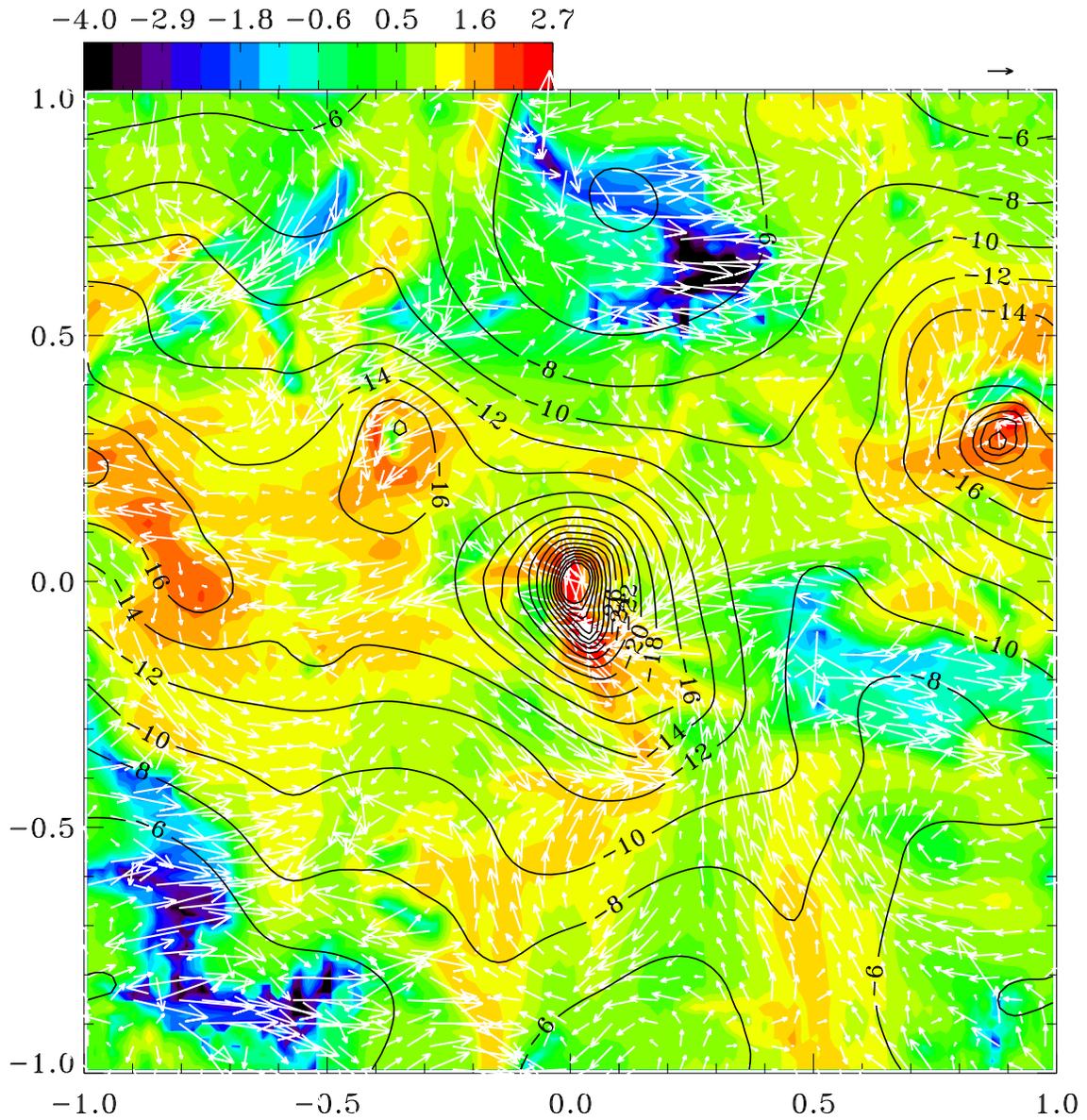}
\caption{Map of the logarithm of the density (in units of $\rho_0$) in 
the ``midplane'' of the condensed sheet. Overlaid are contours of 
gravitational potential (in units of sound speed squared) and velocity 
vectors (with a unit vector of length equal to sound speed shown 
above the panel for comparison). } 
\label{den_pot_vel}
\end{figure}

\begin{figure}
\plotone{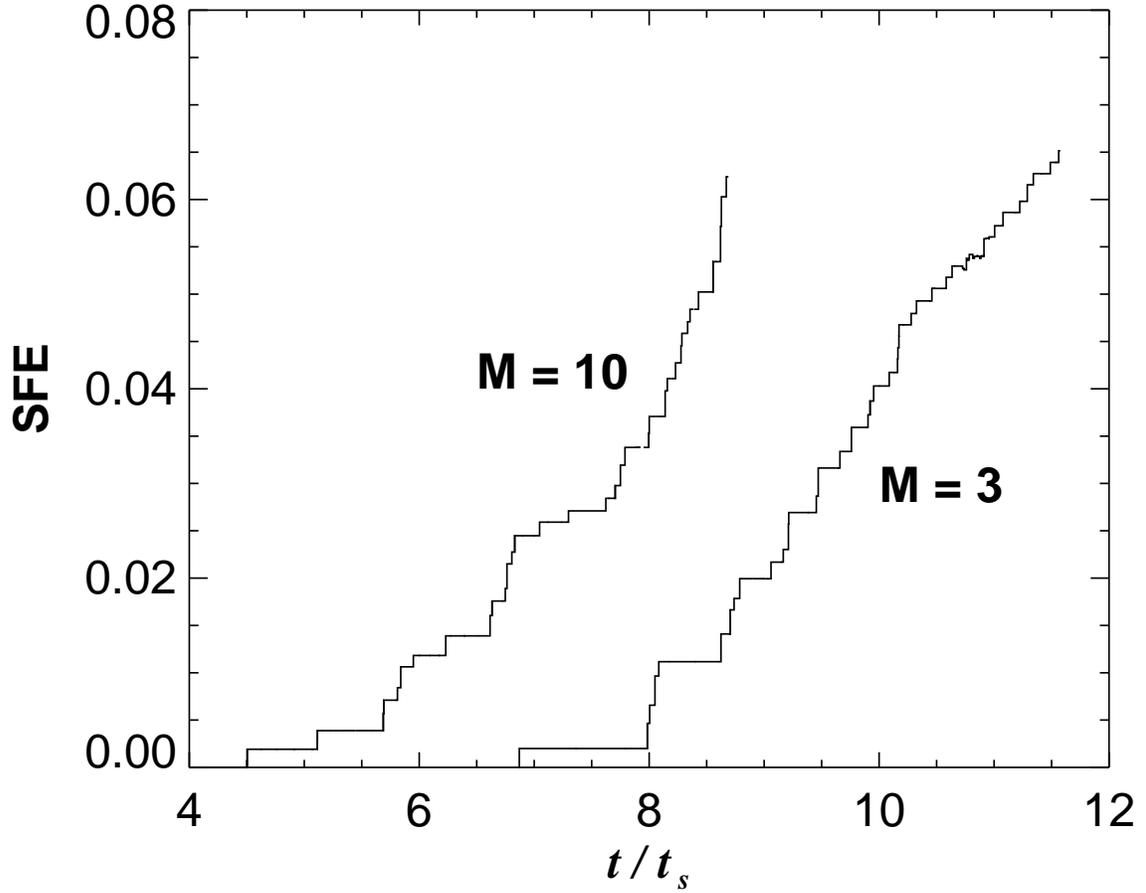}
\caption{Star formation efficiencies as a function of time for two 
models with initial turbulent Mach number ${\cal M}=10$ and $3$ 
(standard model), respectively. Stars form earlier in the initially 
more turbulent cloud, illustrating acceleration of star formation 
by turbulence in strongly magnetized clouds.} 
\label{sfetwo}
\end{figure}

\begin{figure}
\plotone{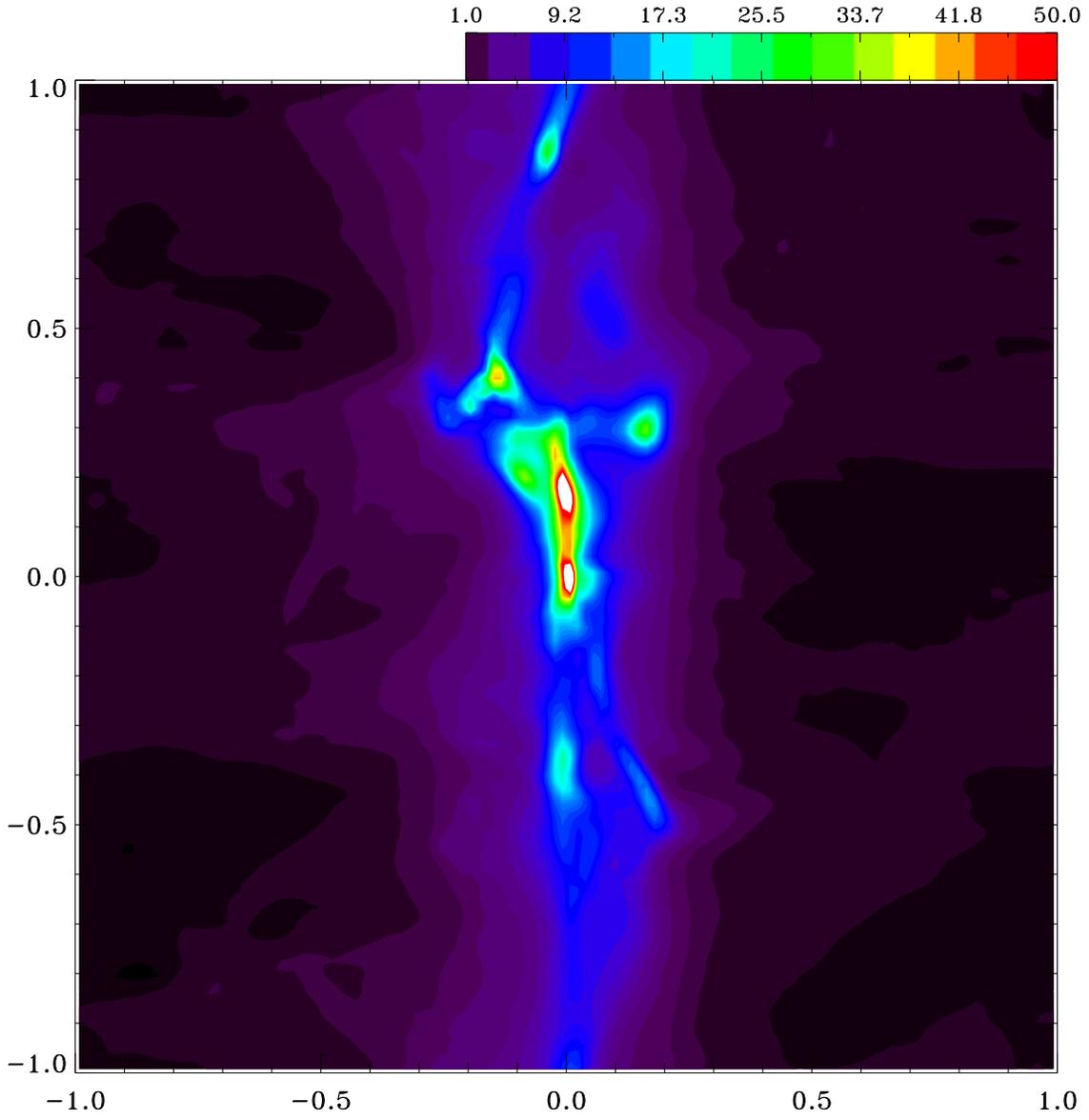}
\caption{Same as Fig.~\ref{colden_y} but for an initial turbulent 
Mach number ${\cal M}=10$, showing that the initially more 
turbulent model has an apparently thicker condensed sheet than 
the standard model.} 
\label{colden_y_M10}
\end{figure}

\begin{figure}
\plotone{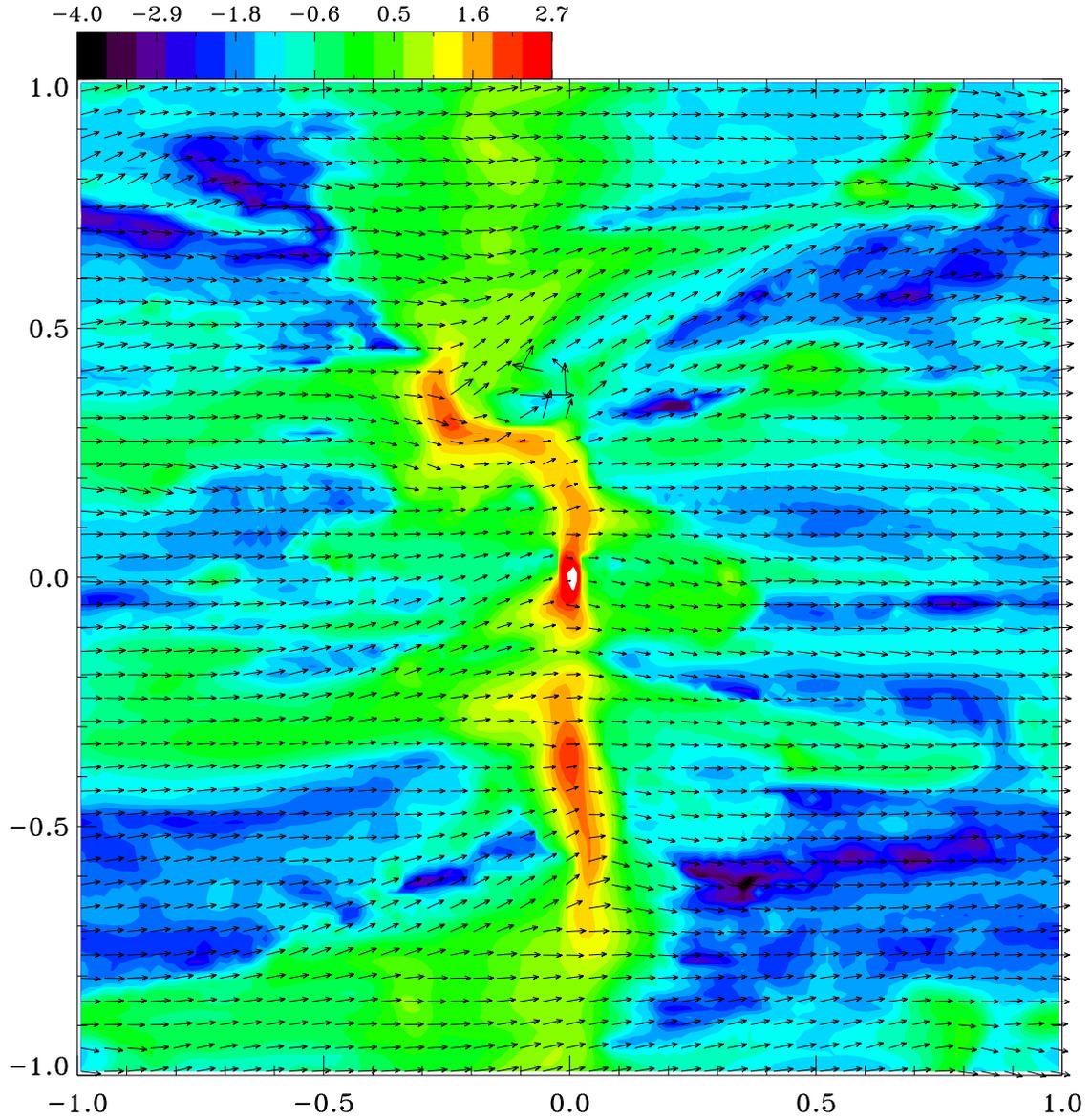}
\caption{Map of the logarithm of the density (in units of $\rho_0$) 
in an $xz$ plane that passes through the minimum of gravitational 
potential. Superposed are unit vectors for magnetic field directions. 
Note the prominent warp at $z\sim 0.3$. Superposition of such warps
along the line of sight can make the condensed sheet appear thicker
than its intrinsic thickness, as illustrated in Fig.~\ref{colden_y_M10}.  
}
\label{den_Bfield_M10}
\end{figure}

\begin{figure}
\plotone{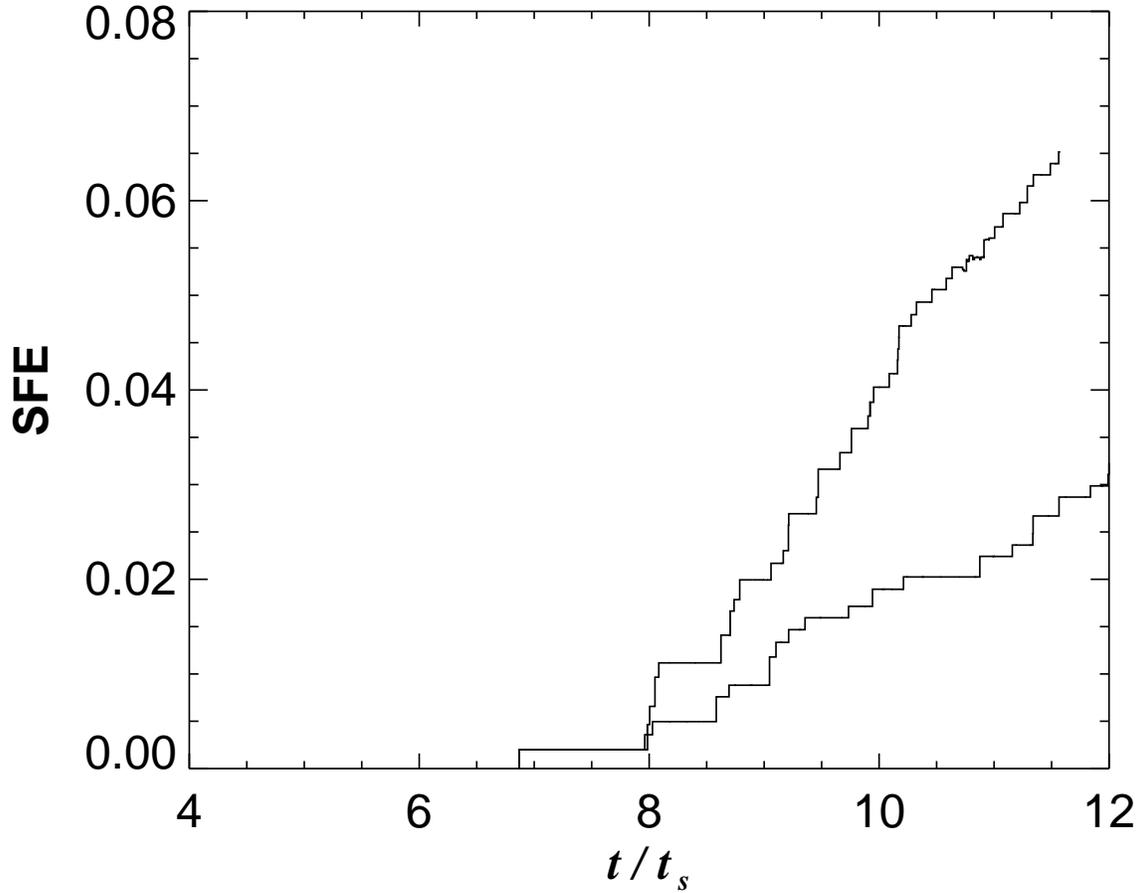}
\caption{Star formation efficiencies as a function of time for two 
models with outflow momentum dominated by, respectively, the 
bipolar jet ({\it upper curve}, standard model) and spherical component 
({\it lower curve}), showing that the spherical component of outflow
is more efficient in slowing down star formation in flattened 
condensations of strongly magnetized clouds. }
\label{sfedependm3}
\end{figure}

\begin{figure}
\plotone{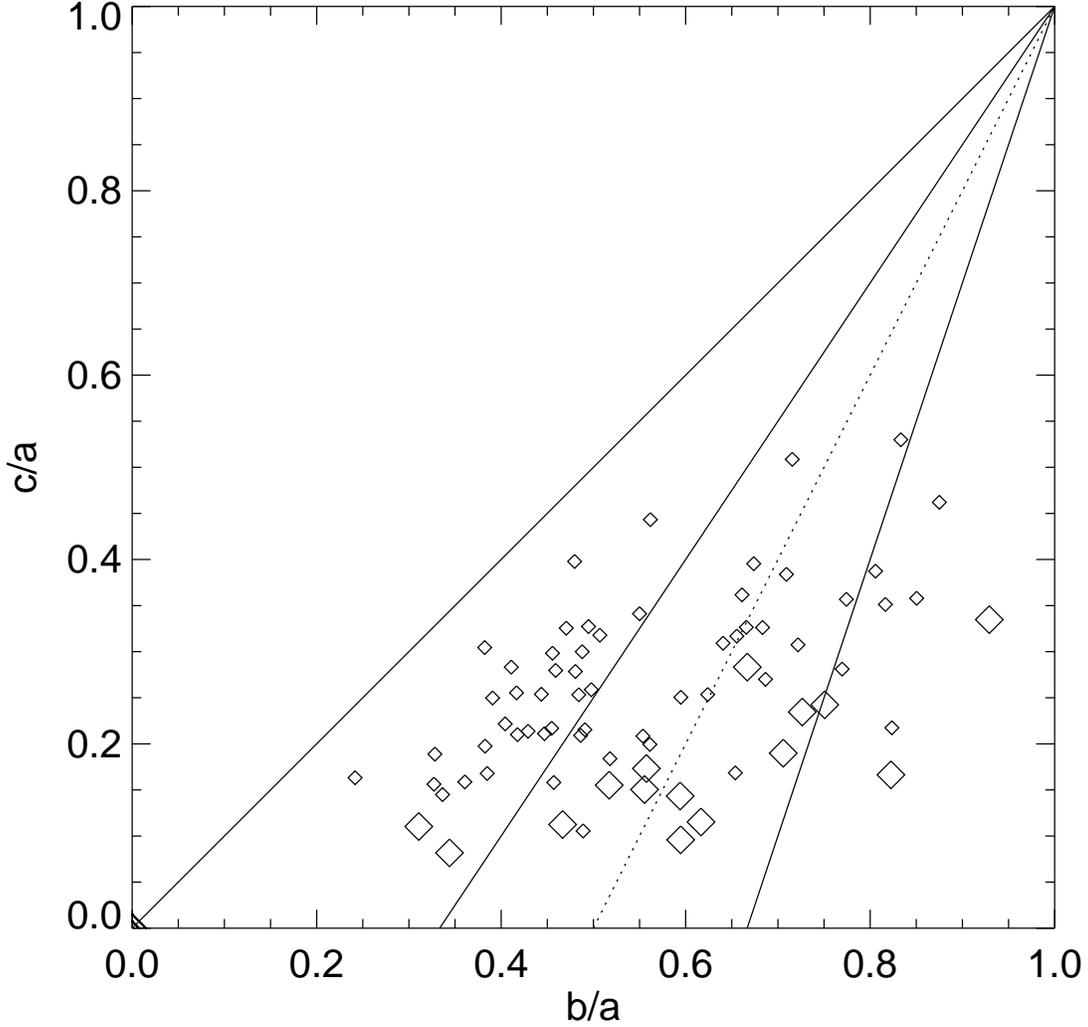}
\caption{Distribution of axis ratios of dense cores formed after the
  first star was created. The abscissa and ordinate are the axis 
ratios of $\xi=b/a$ and $\eta=c/a$, respectively. A large symbol 
indicates a core with a mass greater than $1.5 \left<M_c\right>$, 
while a small symbol indicates a core with a smaller mass. We divide 
the $\xi$-$\eta$ plane into three parts: a prolate group which lies 
above the line connecting $(\xi, \eta)=(1.0, 1.0)$ and (0.33, 0.0), 
an oblate group which lies below the line connecting $(\xi,
\eta)=(1.0, 1.0)$ and (0.67, 0.0), and a triaxial group which is 
everything else. For comparison we draw the dotted line that connects 
$(\xi, \eta)=(1.0,1.0)$ and (0.5, 0.0). 
}
\label{fig:coreshapenew}
\end{figure}

\begin{figure}
\plotone{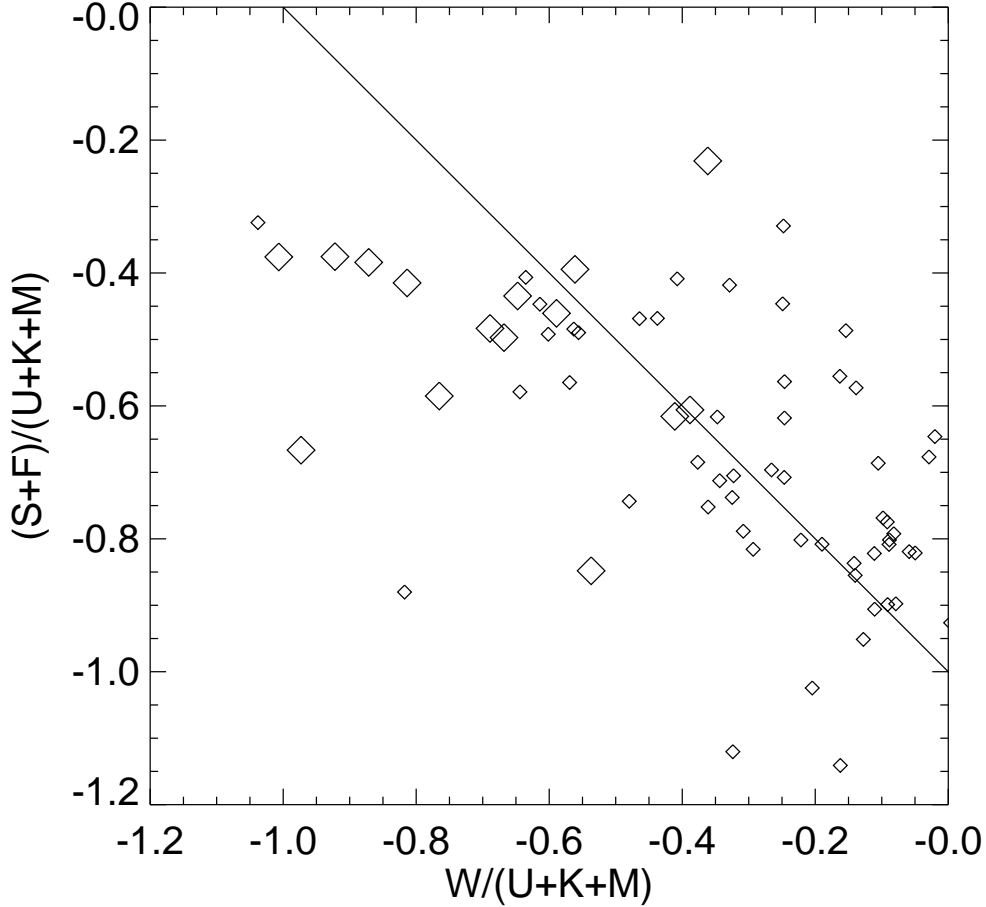}
\caption{The relationship between the sum of the two surface terms, 
$S+F$, and the gravitational term, $W$, in the virial equation. 
They are normalized to the sum of the internal terms, $U+K+M$.
The solid line indicates the virial equilibrium, $U+K+W+S+M+F=0$.
The large and small symbols are the same 
as those of Figure \ref{fig:coreshapenew}.
For the cores that lie below the line, the left hand side 
of the virial equation (\ref{eq:virial}) is negative and 
thus expected to be bound. 
All others that lie above the line are unbound and expected to 
disperse away, if they do not gain more mass through accretion
 and/or merging with other cores, or reduce internal 
support through turbulence dissipation and/or magnetic flux 
reduction through ambipolar diffusion. 
}  
\label{fig:virialparamnew}
\end{figure}

\begin{figure}
%\epsscale{0.6}
\plotone{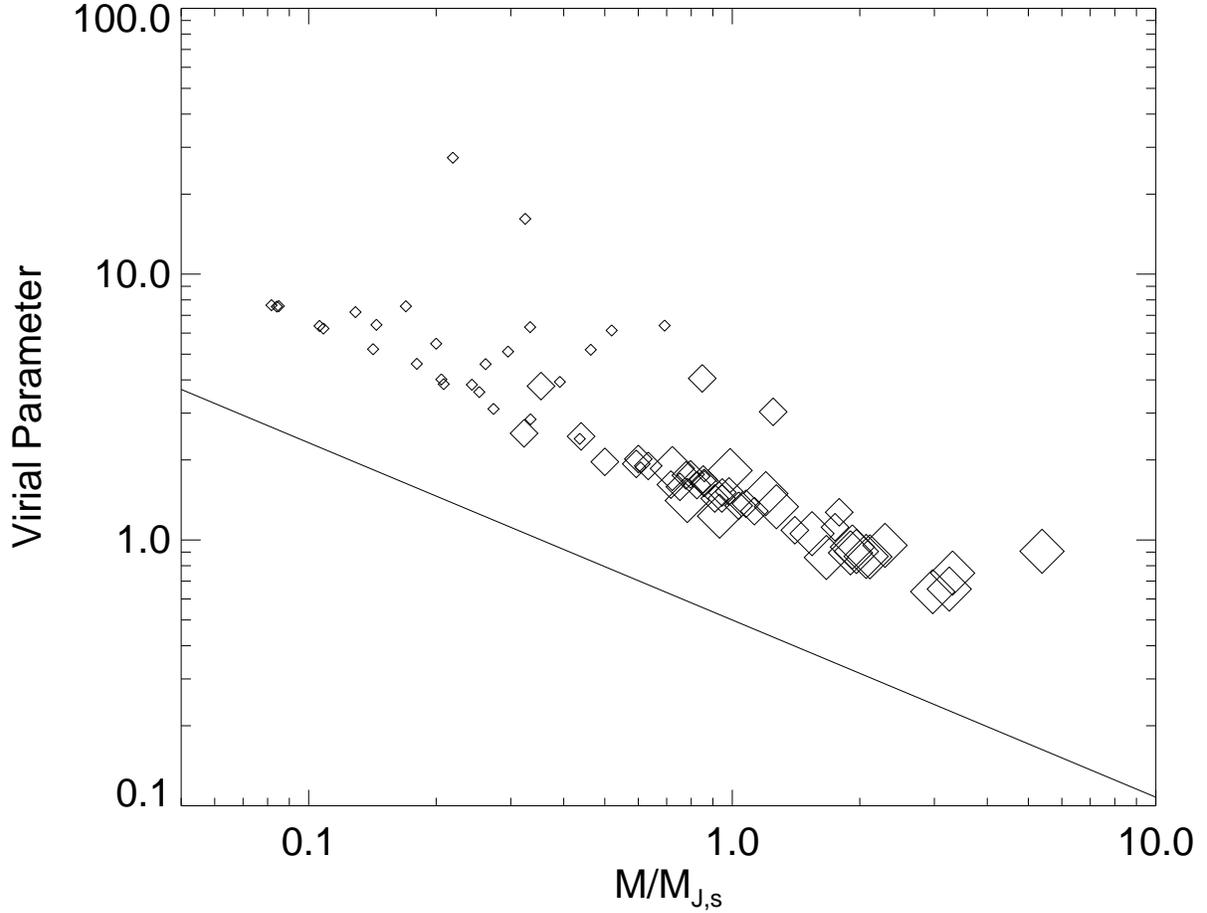}
\caption{The virial parameter as a function of the core mass.
Large, intermediate, and small symbols denote, respectively, 
cores with $\Gamma_c \le 0.5$, $0.5 < \Gamma_c \le 0.7$, 
and $\Gamma_c > 0.7$, where $\Gamma_c$ is the flux-to-mass ratio 
of a core. The virial parameter can be fitted by a power law of
$\alpha_{\rm vir}\propto M_c^{-2/3}$, indicated by the solid line.
}  
\label{fig:virialrationew}
\end{figure}

\begin{figure}
%\epsscale{0.6}
\plotone{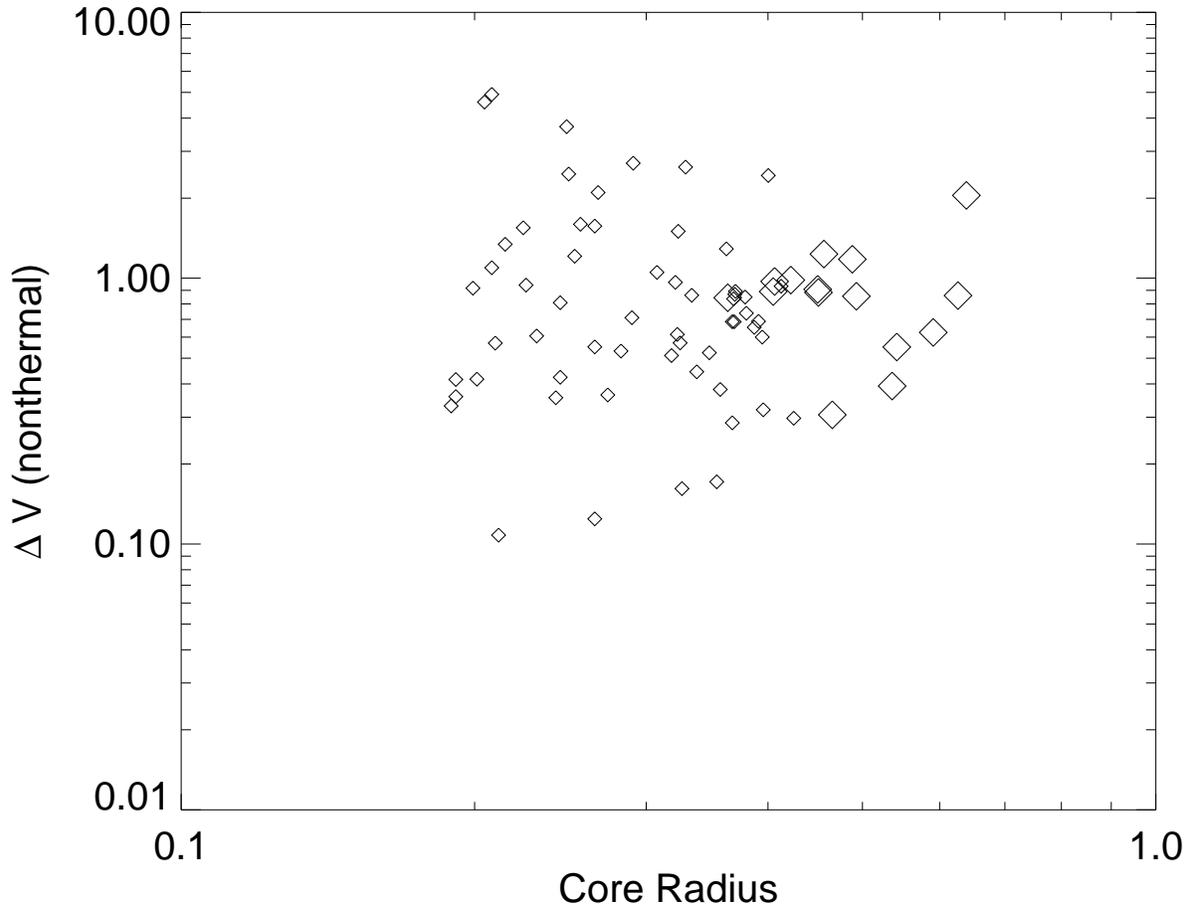}
\caption{
The nonthermal velocity dispersion as a function of core radius.
The velocity dispersion is normalized to the sound speed $c_s$,
and radius to the Jeans length of the Spitzer sheet $L_{s}$.
The large and small symbols are the same as those of 
Figure \ref{fig:coreshapenew}.
There is apparently no correlation between the velocity 
dispersion and the radius, in good agreement with 
the observed linewidth-size relation of the dense cores in 
the Taurus molecular clouds (Onishi et al. 2002). 
} 
\label{fig:velocitysizenew}
\end{figure}

\begin{figure}
%\epsscale{0.6}
\plotone{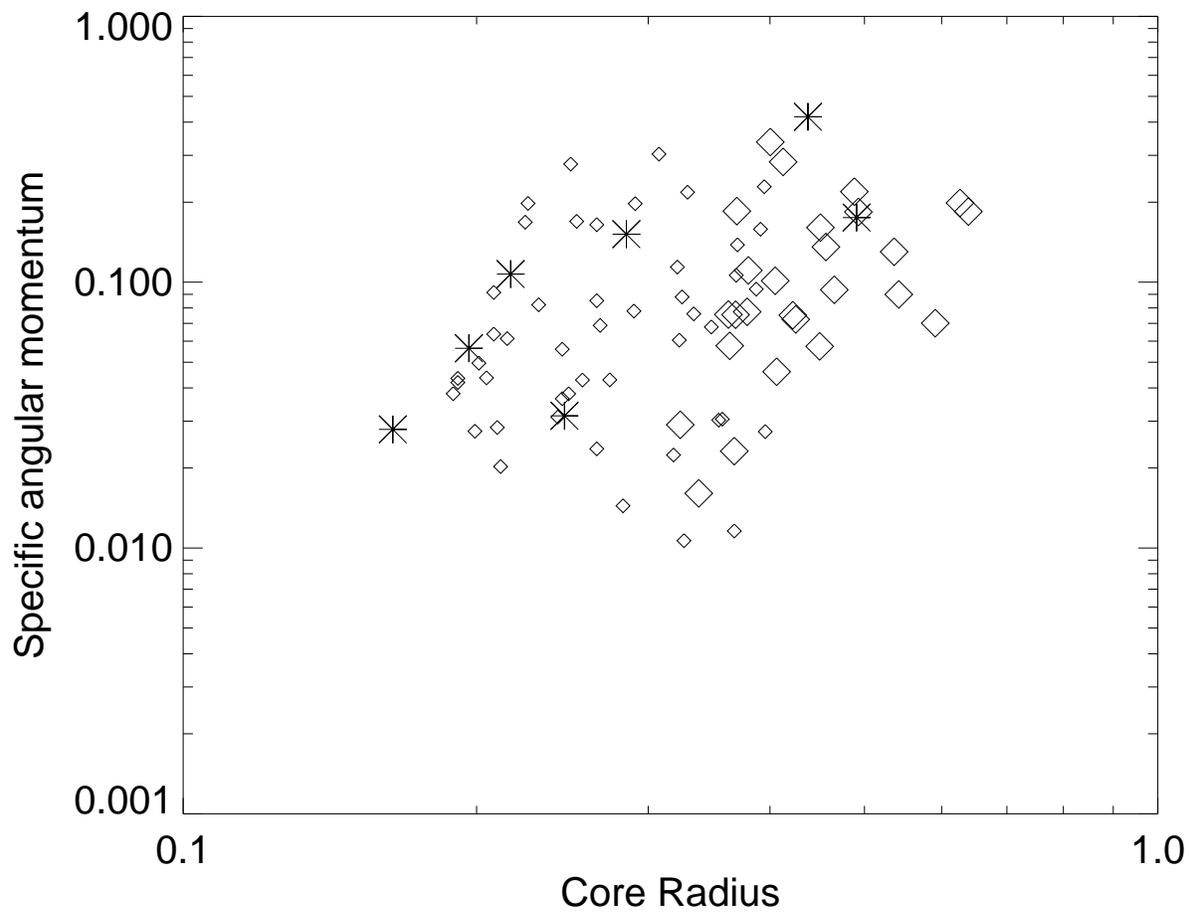}
\caption{Distribution of specific angular momentum as a function of core 
radius. The specific angular momentum is normalized to $c_s L_s = 1.1
\times 10^{22}$ cm$^{2}$ s$^{-1}$ (for our fiducial cloud parameters),
and radius to the Jeans length of the Spitzer sheet $L_s$. 
The large and small symbols are the same as those of Figure 
\ref{fig:coreshapenew}. For comparison, the specific angular momenta 
of 7 starless N$_2$H$^+$ cores are plotted as asterisks (data from Table 5 
of Caselli et al. 2002). 
}  
\label{am1}
\end{figure}

\begin{figure}
%\epsscale{0.6}
\plotone{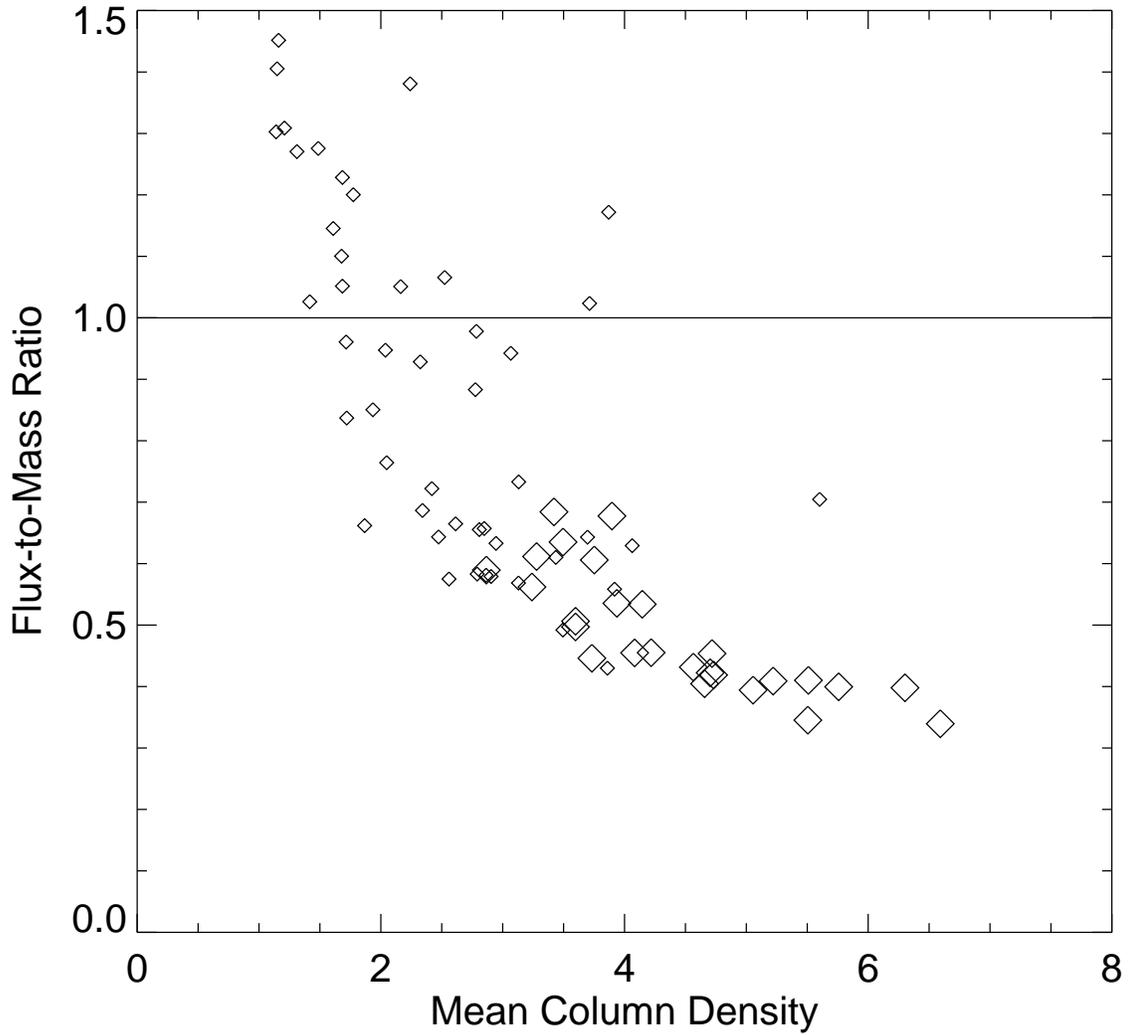}
\caption{The magnetic flux-to-mass ratio as a function of the column
 density of the core.
The large and small symbols are the same 
as those of Figure \ref{fig:coreshapenew}.
The cores below the solid line $\Gamma_c=1$ are 
magnetically supercritical.
The flux-to-mass ratio tends to be smaller for more massive cores.
}  
\label{fig:mass2fluxnew}
\end{figure}

\begin{figure}
%\epsscale{0.4}
\plotone{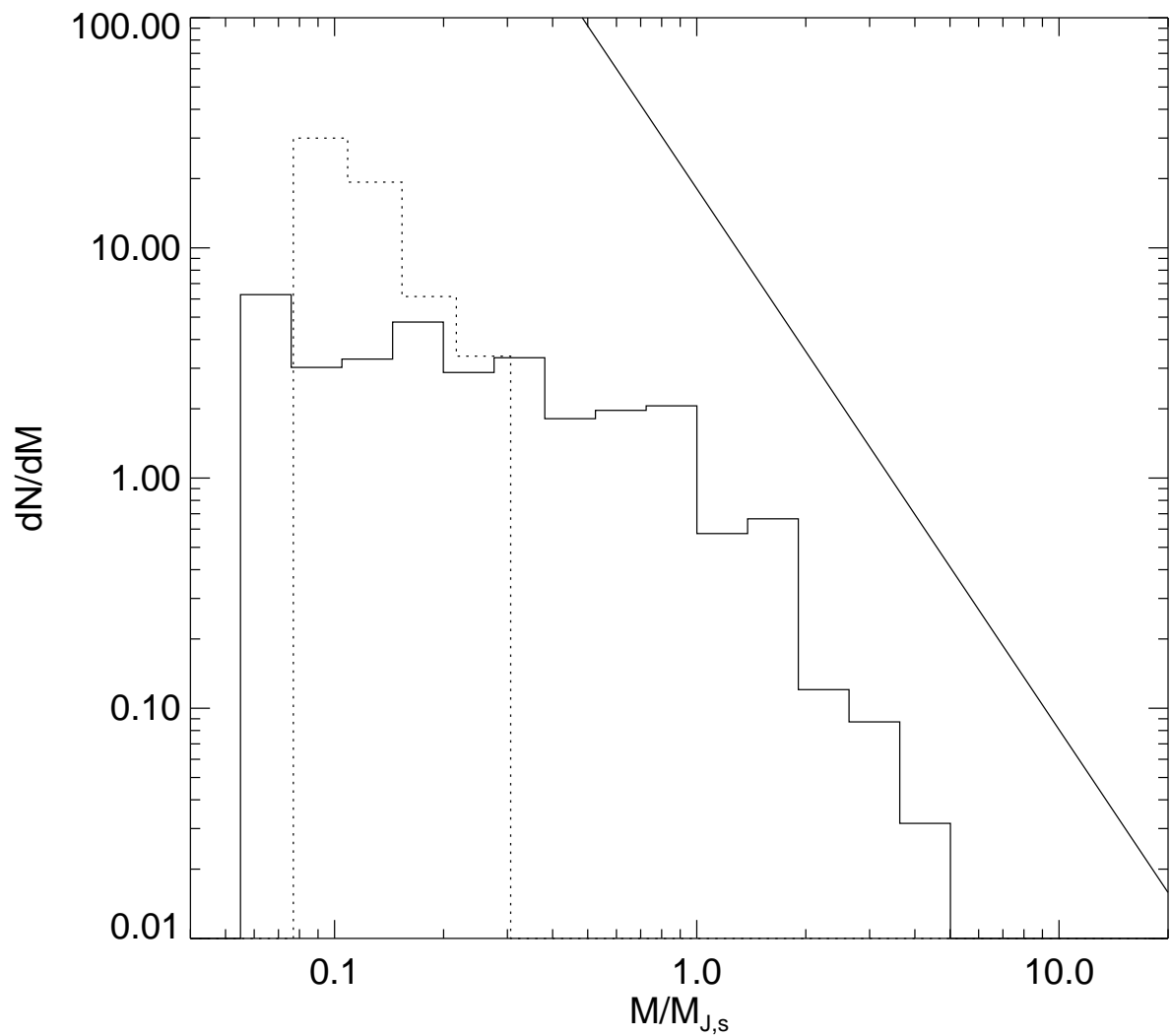}
\caption{Core mass spectrum. The core mass is normalized to 
the Jeans mass of the Spitzer sheet, $M_{J,s}$.
The solid histogram indicates the mass spectrum of dense cores.
For comparison, we show the stellar mass spectrum
with the dotted histogram.
The solid line denotes the power law $dN/dM \propto M^{-2.35}$, 
the Salpeter IMF.
There is a prominent break around $1 M_{J,s}$ in the core mass
 spectrum. Above the break, the spectrum is in good agreement 
 with the Salpeter IMF. 
The stellar mass spectrum can be also fitted by the Salpeter IMF.
}  
\label{fig:coremf}
\end{figure}

We have performed 3D simulations of star formation in turbulent, 
mildly magnetically subcritical clouds including protostellar 
outflows and ambipolar diffusion. The simulations are motivated
by observations of molecular gas and young stars in the Taurus
cloud complex and the realization that cold neutral HI gas, a 
likely progenitor of molecular gas, is magnetically subcritical
in general. They are natural extensions of our previous 2D 
calculations. The main results are:  

1. The decay of turbulence leads to gravitational condensation 
of material along the field lines. In the ideal MHD limit, the 
resulting structure closely resembles a smooth Spitzer sheet, 
as found previously (Krasnopolsky \& Gammie 2005). Ambipolar 
diffusion changes the structure drastically. It creates 
supercritical material that behaves differently from the 
subcritical background in a fundamental way: it can be condensed 
further by self-gravity. As a result, the distributions of mass 
with respect to volume and column densities are greatly broadened. 
Indeed, the probability distribution function (PDF) of the (more 
easily observable) column density can be fitted approximately 
by a broad, lognormal distribution (see Fig.~\ref{colden_PDF}), 
as is the case for weakly magnetized clouds fragmented by strong 
turbulence (e.g., Padoan \& Nordlund 2002; P. S. Li et al. 2004). 
It may be difficult to distinguish the turbulent fragmentation 
in weakly magnetized clouds and the ambipolar diffusion-enabled 
gravitational fragmentation in strongly magnetized clouds based 
on mass distribution. 

\begin{figure}
%\epsscale{0.65}
\plotone{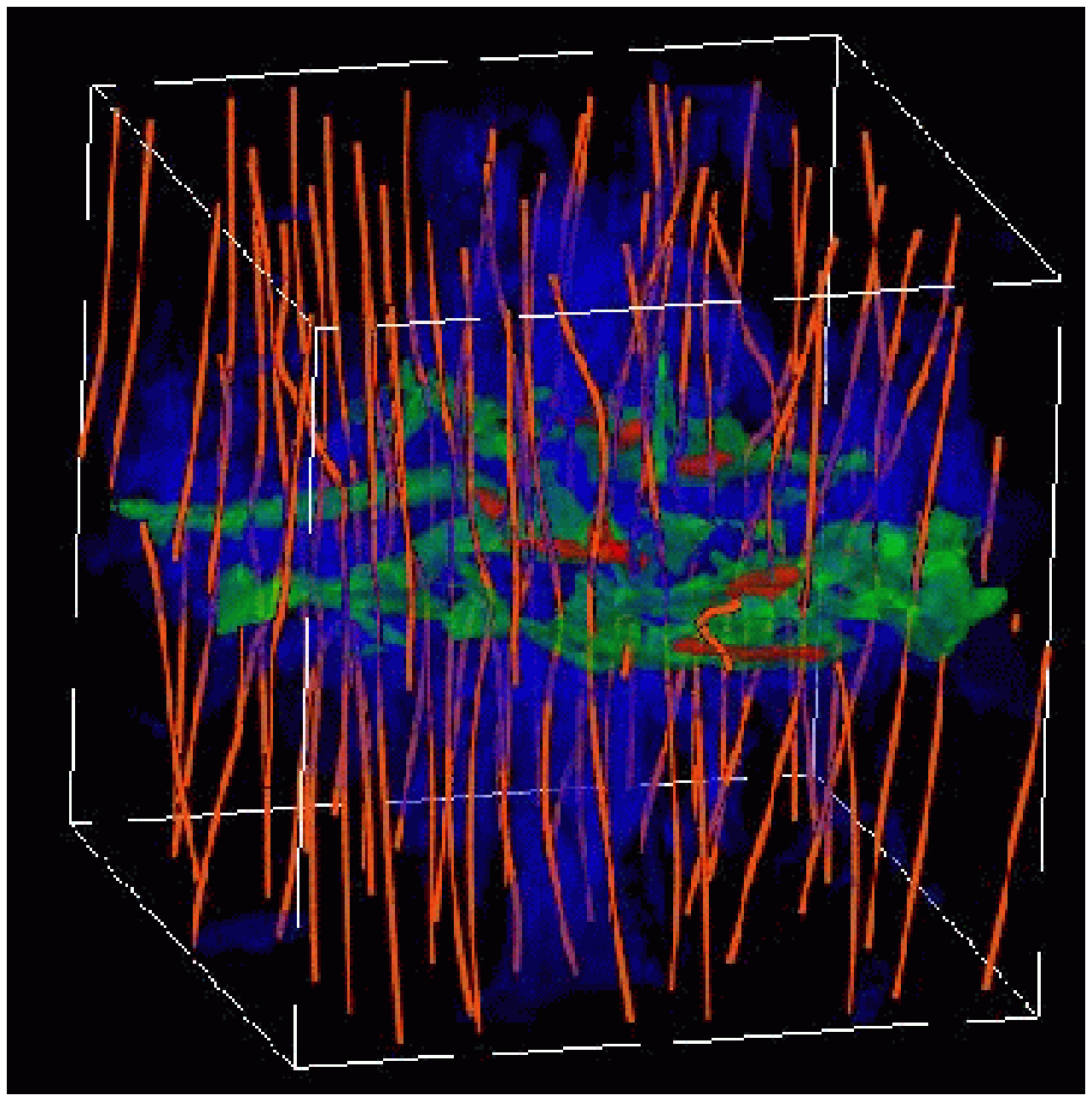}
\caption{3D view of density distribution and field lines of the  
standard model at the same time as in
Figs.~\ref{colden_y} and \ref{colden_critical}. The isodensity surfaces 
have values $\rho=0.5$
({\it blue}), 15 ({\it green}) and 60~$\rho_0$ ({\it red}). They 
represent three  distinct cloud components:
turbulent diffuse halo,  fragmented condensed sheet, and dense cores.  
The sheet is nearly magnetically
critical,  whereas the halo and cores are significantly subcritical and 
supercritical, respectively.}
\label{3D}
\end{figure}

2. The magnetized clouds evolve into a configuration strongly 
stratified in density at late times, with dense cores embedded 
in fragmentary condensations which, in turn, are surrounded 
by a more diffuse halo (see Fig.~\ref{3D}). 
The condensed material can be significantly warped and may appear 
thicker than its intrinsic width, even when viewed edge-on, 
because of superposition of warps along the line of sight. While 
the diffuse halo is highly turbulent, the condensations are 
more quiescent, with moderately supersonic velocity dispersions
in general. One reason is that the bulk of the condensed material 
is magnetically supported, which allows more time for the 
turbulence to decay. Another is their relatively low Alfv\'en 
speed (comparable to the sound speed), which forces highly 
supersonic motions to dissipate quickly, through shock 
formation. The moderately supersonic turbulence is maintained, 
however, for many local free-fall times, through a combination 
of outflow feedback, MHD waves, and gravitational motions 
induced by ambipolar diffusion.
%, even in the absence of any external driving. 
  
3. Stars form at a low rate. Only one percent or less of the cloud 
material is converted into stars in a free-fall time at the 
characteristic density of the condensed material. The slow star 
formation cannot be due to support by turbulence because the  
star-forming, condensed material is rather quiescent. It is 
regulated by magnetic fields. The low rate of star formation is 
not sensitive to the level of initial turbulence, although more 
turbulent clouds produce stars early, consistent with the picture 
of turbulence-accelerated star formation in strongly magnetized 
clouds. The rate of star formation is further reduced by 
protostellar outflow, particularly the component in the plane 
of mass condensation.   

4. Dense cores formed in our simulations are typically triaxial.  
More massive cores tend to have lower magnetic flux-to-mass
ratios, be more tightly bound gravitationally, and be more 
oblate. Virial analysis indicates that the surface term due to 
external kinetic motions contributes significantly in general, 
especially for less massive cores. We find a larger number 
of bound cores than in previous ideal MHD simulations even 
though our clouds are more strongly magnetized; they are made 
possible by ambipolar diffusion. The cores have specific 
angular momenta comparable to the observed values, and a mass 
spectrum that resembles the Salpeter stellar IMF towards the 
high mass end. The spectrum flattens near the characteristic 
Jeans mass of the condensed material, which is a few solar 
masses for typical parameters. The stellar mass spectrum also 
resembles the Salpeter IMF. 

5. The Taurus molecular cloud complex may represent the best case 
for magnetically regulated star formation. The strongest evidence 
comes from the existence of a large amount of dense, relatively 
quiescent, C$^{18}$O gas, which produces stars at a rate two orders 
of magnitude below the maximum free-fall rate. Most likely, the 
star formation is regulated by a strong, ordered magnetic field. 
Several lines of evidence suggest that a magnetic field of the 
required strength (of order $\sim 20 \mu G$) is indeed present 
in the region, although direct Zeeman measurements are available 
for only three cores, and their interpretations are complicated 
by projection effects. 

6. We speculate that, depending on the (uncertain) rates of 
turbulence replenishment and outflow feedback, there may be 
three distinct modes of star formation in strongly magnetized 
clouds: (a) inefficient star formation through 
turbulence-accelerated ambipolar diffusion at dispersed 
locations in relatively diffuse, magnetically subcritical 
clouds, (b) more coherent, but still inefficient, 
magnetically regulated star formation in relatively 
quiescent, nearly magnetically critical condensations, 
(c) more efficient cluster formation in magnetically 
supercritical clumps that may be regulated by outflow-driven 
protostellar turbulence. Much work remains to be done to 
understand these modes and their possible interconnection. 

\acknowledgments 
This work is supported in part by a Grant-in-Aid for Scientific Research 
of Japan (18540234, 20540228),  
and NSF and NASA grants (AST-0307368 and NAG5-12102). 
Parts of this work were performed while the authors were in residence
at the Kavli Institute for Theoretical Physics at UCSB.
The numerical calculations were carried out mainly 
on NEC SX8 at YITP in Kyoto University, 
on Fujitsu VPP5000 at the Center for Computational Astrophysics, CfCA, 
of the National Astronomical Observatory of Japan,
and on Hitachi SR8000 at Center for Computational Science 
in University of Tsukuba.

\begin{deluxetable}{lllll}
%\tabletypesize{\scriptsize}
%\rotate
\tablecolumns{4}
\tablecaption{Model Parameters}
\tablewidth{\columnwidth}
\tablehead{
 \colhead{Model}     & \colhead{${\cal M}$} & \colhead{$\eta$} 
& \colhead{AD}  & \colhead{Note} 
}
\startdata
S0   & 3 & 0.75 & yes & standard model\\
S1   & 3 & 0.75 & yes & different realization of initial velocity field\\
S2   & 3 & 0.25 & yes & stronger spherical outflow component\\
M0   & 10 & 0.75 & yes   & stronger initial turbulence \\
I0   & 3 & N/A & no   & ideal MHD counterpart of standard model\\
\enddata
\tablecomments{
All models have the same initial distributions of mass and magnetic 
field. The field strength is given by the dimensionless ratio of 
magnetic to thermal pressure $\alpha=24$, corresponding to a 
mildly magnetically subcritical cloud of flux-to-mass ratio 
$\Gamma_0=1.1$. The models differ in initial turbulent Mach number 
${\cal M}$, jet momentum fraction $\eta$, random realization 
of the initial velocity field, and whether ambipolar diffusion 
(AD) is included or not. 
}
\label{tab:model parameter}
\end{deluxetable}

\clearpage

\clearpage

\clearpage

\begin{thebibliography}{99}
\bibitem[]{}
Alves, F. O. \& Franco, G. A. P. 2007, \aap, 470, 597
\bibitem[]{}
Alves, J., Lombardi, M. \& Lada, C. F. 2007, \aap, 462, 17
\bibitem[]{}
Andre, P., Ward-Thompson, D. \& Barsony, M. 1993, \apj, 406, 122
\bibitem[]{}
Ballesteros-Paredes, J. Hartmann, L., \& Vazquez-Semadeni, E. 1999,
          \apj, 527, 285
\bibitem[]{}
Basu, S. \& Ciolek, G. E., 2004, \apj, 607, 39
\bibitem[]{}
Bertoldi, F. \& McKee, C. F. 1992, \apj, 395, 140
\bibitem[]{}
Caselli, P. \& Myers, P. C. 1995, \apj, 446, 665
\bibitem[]{}
Dib, S., Vazquez-Semadeni, E., Kim, J., Burkert, A., \& Shadmehri,
          M. 2007, \apj, 661, 262 
\bibitem[]{}
Elmegreen, B. G. 2007, \apj, 668, 1064
\bibitem[]{}
Gammie, C. F., Lin, Y.-T., Stone, J. M., \& Ostriker, E. C. 2003, \apj,
          592, 203
\bibitem[]{}
Goldsmith et al. 2008 (astro-ph/arXiv:0802.2206)
\bibitem[]{}
Gomez, M. Hartmann, L., Kenyon, S. J., \& Hewett, R. 1993, \apj, 105, 1927
\bibitem[]{}
  Hartmann, L., Ballesterios-Paredes, J. \& Bergin, E. A. 2001, \apj,
          562, 852
\bibitem[]{}
Heiles, C. \& Crutcher, R. M. 2005, in Cosmic Magnetic Fields, Lecture
Notes in Physics, vol. 664, pp137-182
\bibitem[]{}
Heiles, C. \& Troland, T. H. 2005, \apj, 624, 773
\bibitem[]{}
 Heyer, M., Gong, H., Ostriker, E. Brunt, C. 2008, (astro-ph/arXiv:0802.2084)
\bibitem[]{}
Heyer, M. H., Frederick, J., Snell, R. L., Schloerb, F. P., Strom,
          S. E., Goldsmith, P. F., \& Strom, K. M. 1987, \apj, 321, 855 
\bibitem[]{}
Ikeda, N., Sunada, K., \& Kitamura, Y. 2007, \apj, 665, 1194
\bibitem[]{}
Kazes, I. \& Crutcher, R. M. 1986, \aap, 164, 328
\bibitem[]{}
Kenyon, S. J. \& Hartmann, L. 1995, ApJS, 101, 117
\bibitem[]{}
Kenyon et al. 2008, to appear in Handbook of Star Forming Regions, ASP 
Conference Series, ed. B. Reipurth
\bibitem[]{}
  Krasnopolsky, R. \& Gammie, C. F. 2005, \apj, 635, 1126
\bibitem[]{}
Krumholz, M. R. \& Tan, J. C. 2007, \apj, 654, 304
\bibitem[]{}
        Krumhotz, M. \& McKee, C. F. 2005, ApJ, 630, 250
\bibitem[]{}
   Kudoh, T., Basu, S., Ogata, Y., \& Yabe, T. 2007, \mnras, 380, 499
\bibitem[]{}
   Kudoh, T., \& Basu, S. 2003, \apj, 595, 842
\bibitem[]{}
   Kudoh, T., \& Basu, S. 2006, \apj, 642, 270
\bibitem[]{}
Lada, E. A., Bally, J., \& Stark, A. A., 1991, \apj, 368, 432
\bibitem[]{}
Lada, C. J., Muench, A. A>, Rathborne, J., Alves, J.F., Lombardi,
          M. 2008, \apj, 672, 410
\bibitem[]{}
Larson, R. B. 1981, \mnras, 194, 809 
\bibitem[]{}
  Li, P. S., Norman, M. L., Mac Low, M.-M. \& Heitsch, F. \apj, 605, 800
\bibitem[]{}
       Li, Z.-Y. \& Nakamura, F. 2004, \apj, 609, L83  
\bibitem[]{}
       Li, Z.-Y. \& Nakamura, F. 2006, \apj, 640, L187  
\bibitem[]{}
        Mac Low, M.-M., Klessen, R. S., Burkert, A., \& Smith, M. D. 
1998, \prl, 80, 2754
\bibitem[]{}
        Matzner, C. D. 2007, \apj, 659, 1394
\bibitem[]{}
        Matzner, C. D. \& McKee, C. F. 2000, ApJ, 545, 364
\bibitem[]{412}
   McClure-Griffiths, N. M., Dickey, J. M., Gaensler, B. M., Green,
          A. J., \& Haverkorn, M. 2006, \apj, 652, 1339
\bibitem[]{}
        McKee, C. F. 1989, \apj, 345, 782 
\bibitem[]{}
  McKee, C. F. \& Ostriker, E. C. 2007, \araa, 45, 565
\bibitem[]{}
McKee, C. F. \& Zweibel, E. G. 1992, \apj, 399, 551
\bibitem[]{}
      Mouschovias, T.  \& Ciolek, G. 1999,
   in The Origins of Stars and Planetary Systems, ed. C. Lada \&
   N. Kylafis (Kluwer), p. 305
\bibitem[]{}
Muench, A. A. Lada, C. J., Rathborne, J. M. Alves, J. F., \& Lombardi,
          M. 2007, \apj, 671, 1820
\bibitem[]{}
Myers, P. C. \& Lazarian, A. 1998, \apj, 507, 157
\bibitem[]{}
   Nakamura, F. \& Li, Z.-Y. 2005, \apj, 631, 411
\bibitem[]{}
   Nakamura, F. \& Li, Z.-Y. 2007, \apj, 662, 395
\bibitem[]{}
Nakamura, F., Matsumoto, T., Hanawa, T. \& Tomisaka, K. 1999, \apj,
     510, 274 
\bibitem[]{}
   Nakano, T. 1984, Fundam. Cosmic. Phys., 9, 139
\bibitem[]{}
        Nakano, T. \& Nakamura, T. 1978, PASJ, 30, 681
\bibitem[]{}
Novak, G., Dotson, J. L. \& Li, H. 2007, \apj, in print (arXiv:0707.2818)
\bibitem[]{}
    Onishi, T., Mizuno, A., Kawamura, A., Ogawa, H. \& Fukui, Y. 
       1996, \apj, 465, 815
\bibitem[]{}
    Onishi, T., Mizuno, A., Kawamura, A., Tachihara, K. \& Fukui, Y. 
       2002, \apj, 575, 950
\bibitem[]{}
        Ostriker, E. C., Gammie, C. F. \& Stone, J. M. 2001, \apj, 546, 980
\bibitem[]{}
       Padoan, P. \& Nordlund, A. 2002, \apj, 576, 870
\bibitem[]{}
Palla, F. \& Stahler, S. W. 2000, \apj, 540, 255
\bibitem[]{}
Palla, F. \& Stahler, S. W. 2002, \apj, 581, 1194
\bibitem[]{}
       Price, D. \& Bate, M. R. 2008, \mnras, in press
\bibitem[]{}
        Shu, F. H. 1991, Physics of Astrophysics, Vol. 2: Gas Dynamics
          (Mill Valley: University Science Books)
\bibitem[]{}
        Shu, F. H., Adams, F. C., \& Lizano, S. 1987, \araa, 25, 23
\bibitem[]{}
        Shu, F. H. \& Li, Z.-Y. 1997, \apj, 475, 251
\bibitem[]{}
Shu, F. H., Galli, D., Lizano, S., Glassgold, A. E. \& Diamond,
P. H. 2007, \apj, 665, 535
\bibitem[]{}
Tamura, M., Nagata, T., Sato, S. \& Tanaka, M. 1987, \mnras, 224, 413
\bibitem[]{}
        Tilley, D. \& Pudritz, R. 2007, \mnras, 382, 73
\bibitem[]{}
Troland \& Crutcher 2008 (astro-ph/arXiv:0802.2253v1)
\bibitem[]{}
van Dishoeck, E. F. \& Black,J. H. 1988 , \apj, 334, 771
\bibitem[]{}
Williams, J. P., de Geus, E. J., \& Blitz, L. 1994, \apj, 428, 693
\bibitem[]{}
Wolleben, M. \& Reich, W. 2004, \aap, 427, 537
\end{thebibliography}
\end{document}